\documentclass[english,english,aps,prd,amsmath,amssymb,superscriptaddress]{revtex4}
\usepackage[T1]{fontenc}
\usepackage[latin9]{inputenc}
\usepackage{amsmath}
\usepackage{amssymb}
\usepackage{esint}

\makeatletter

\providecommand{\tabularnewline}{\\}

\@ifundefined{textcolor}{}
{%
 \definecolor{BLACK}{gray}{0}
 \definecolor{WHITE}{gray}{1}
 \definecolor{RED}{rgb}{1,0,0}
 \definecolor{GREEN}{rgb}{0,1,0}
 \definecolor{BLUE}{rgb}{0,0,1}
 \definecolor{CYAN}{cmyk}{1,0,0,0}
 \definecolor{MAGENTA}{cmyk}{0,1,0,0}
 \definecolor{YELLOW}{cmyk}{0,0,1,0}
 }

\usepackage{slashed}
\usepackage{babel}
\usepackage{bbm}

\makeatother

\usepackage{babel}

\begin{document}

\title{Gauge Systems with Finite Chemical Potential in 2+1 Dimensions by
Bosonization}

\author{M.J.Luo}

\affiliation{Department of Physics, Tsinghua University, Beijing 100084, People's
Republic of China}
\begin{abstract}
We present a bosonization method to study generic low energy behavior
of gauge systems with finite chemical potential in 2+1 dimensions.
Benefit from the existence of gap (e.g. Gribov gap) in gauge systems
at low energy, the fermion fields can be explicitly bosonized by new
gauge fields. When chemical potential of the gauge systems is introduced,
we find that topological terms (such as Chern-Simons term in 2+1D)
as constraints inevitably emerge at low energy. The fermion sign problem
at finite chemical potential and its deep connection to the Chern-Simons
theories are discussed. The Wilson's criteria of confinement in pure
gauge theories is generalized to finite chemical potential case. The
chemical potential dependence of physical quantities at strong coupling
are explicitly calculated, including the expectation value of the
Wilson loop, the confining potential and the confined/deconfined transition
temperature. The bosonization puts discussions on chiral symmetry
breaking and confined/deconfined transition on an equal footing, so
it is suitable for the study of the subtle interplay between them.
We find that the chiral symmetry breaking is a necessary (not sufficient)
condition for the confinement in 2+1D, and argue that the confined/deconfined
phases are not characterized by any local symmetries but distinguished
by their non-local topologies. The low energy modes of the strongly
coupled gauge systems in a non-symmetry breaking phase is also discussed.
The results of the paper can be widely applied to real strongly coupled
gauge systems, e.g. high-temperature superconductor and quantum chromodynamical
systems in 2+1 dimensions.
\end{abstract}
\maketitle

\section{Introduction}

Gauge system is the most fundamental model of our world, not only
describing the basic interactions but also emergent structures in
many body systems in condensed matter. However, when the gauge system
is strongly coupled, standard perturbative methods fails. At present,
there are no systematic , non-perturbative and analytic approaches
to study these systems. In many cases, we rely on numerical simulations,
e.g. quantum Monte Carlo (QMC) on lattice. The QMC on lattice is the
only systematic method with sufficient numerical accuracy in calculating
many thermodynamics quantities of strongly coupled pure gauge system,
such as the quantum chromodynamics (QCD) system without dynamical
quarks. However, the numerical method has its own limitations, the
QMC has the notorious fermion sign problem \cite{PhysRevB.41.9301,PhysRevLett.94.170201}
when the pure gauge system are doped by fermions, or equivalently,
the gauge system is at finite chemical potential. The problem makes
the studying of properties of gauge systems at finite chemical potential
inaccessible.

On the other hand, Landau's theory of phase transition is developed
for system at finite temperature while quantum effects are ignored.
The finite chemical potential induced new phenomenon and phase transition
are in essential due to quantum effects, which are conjectured has
richer structures \cite{Wen:2002zz,2002qgph.conf..233S} than the
thermal (classical) phase transition. It seems very likely that new
states of matter we have not encountered before would emerge in finite
chemical potential region. The phase structure of gauge systems such
as the QCD system \cite{Barbour:1997ej,Philipsen:2008zz} and certain
condensed matter systems \cite{2008RPPh...71a2501L,RevModPhys.78.17}
(e.g. Hubbard model at half-filling and Heisenberg spin system are
established as gauge symmetric systems \cite{PhysRevB.38.745,PhysRevB.38.2926})
at finite chemical potential is one of the most important and practical
issues of quantum matter. Generic phase structures are qualitative
common in these gauge system, e.g. the confined/deconfined phase transition,
chiral symmetry breaking/restoring, high temperature superconducting
phase transition are expected occur in the finite chemical potential
region. It is gradually become a general believe that a theory of
gauge system at finite chemical potential holds the key to understand
the exotic new states of matter observed at finite chemical potential,
e.g. the quark-gluon plasma in the QCD system \cite{PhysRevLett.106.192301,Muller:2007rs},
the strange metal \cite{2009Sci...323..603C} and/or pseudo-gap \cite{1999RPPh...62...61T}
in cuprates. The central question is how these phenomena take place
as a function of chemical potential in these gauge systems.

We suggest that the strategy to overcome the difficulty accessing
the finite chemical potential region is to reformulate the systems
in a more proper language. True enough, the fermionic and bosonic
languages are just mathematical machineries that invented independently
for their special use and for convenient in different realms. The
idea of the paper is very transparent, if we can map a bosonic equivalent
of a theory of fermions, then the gauge system at finite chemical
potential coupling with fermion matter is able to identify with a
new pure gauge system. This line of thought leads to the idea of bosonization
\cite{2004bscs.book.....G}. The problems aroused in the introducing
fermions are hence translated to problems of the well studied theories
of pure gauge system. In 1+1 dimensions, the bosonization has been
shown as a powerful technique in studying low energy behaviors of
many strong correlated condensed matter systems \cite{1981JPhC...14.2585H,Witten:148673}.
A lot of efforts have been spent in generalizing this method to higher
dimensions, the explicit results are seem available in 2+1 dimensions.
The bosonization in 2+1 dimensions has many discussions in literature
e.g. \cite{Marino:215651,1995NuPhB.446..144B,1996PhLB..375..267B,le1997non}.
However, these bosonization only applies to systems without chemical
potential. In this paper, we present a new bosonization rule through
a direct variable replacement which relates the massless fermion theories
with the Chern-Simons theories. The Chern-Simons term is relevant
term controlling the low energy behavior and the higher order contributions
from dynamical Maxwell/Yang-Mills terms are considered as perturbations.
Although the idea of bosonization is not new, the deep implications
and consequences of such recipe to the problem of finite chemical
potential of gauge systems has not been fully studied and investigated.
Through the bosonization recipe there are several focuses in the paper,
(i) how the bosonization impacts on the fermion sign problem at finite
chemical potential, (ii) what the bosonization in general tells us
about the confinement of gauge systems at finite chemical potential,
(iii) what is the relation between the chiral symmetry breaking and
the confinement, and (iv) what is the low energy modes of gauge systems
in a phase without symmetry breaking at finite chemical potential. 

The rest of the paper is organized as follows. We begin by the explicit
formulation of the bosonization recipe and give dictionaries to the
translation of terminologies between fermions and bosons. The completely
bosonized actions for the Abelian and non-Abelian gauge theories are
obtained in the section II. In section III, we discuss the fermion
sign problem by the new language of the bosonization instead of fermions,
and show its deep connection to the induced Chern-Simons theories.
Based on a large energy gap of the effective gauge fields at strong
coupling, we introduce a factorization to alleviate the sign problem.
In section IV, we generalize the Wilson's criteria of confinement
in pure gauge theories to finite chemical potential. The chemical
potential dependence of some physical quantities, including the expectation
value of the Wilson loop, the confining potential and the confined/deconfined
transition temperature, are explicitly calculated at strong coupling
limit. In section V, we discuss the interplay between the chiral symmetry
breaking and the confinement, based on the bosonized formalism of
the chiral order parameter. The low energy modes of gauge systems
in a non-symmetry breaking phase at finite chemical potential is discussed
in section VI. We finally draw conclusions of the paper in section
VI. In the last part of the paper, several appendices are given to
proof the translation of terminologies between the fermionic and its
bosonized languages.

\section{Bosonization in 2+1 Dimensions}

The non-locality nature of fermion is largely responsible for the
fermion sign problem. The basic idea of bosonization is that a non-local
bosonic string encodes all relevant information of a fermion. In 1+1
dimension, bosonic scalar field is enough for the bosonization. But
in 2+1 dimensions, the description of the non-local bosonic string
requires at least vector fields. Following the bosonic-string-end-point
picture of fermion, a fermion located at point $\vec{x}$ at given
time $\tau$ is identical to a phase string connecting a spatial fixed
point $\vec{x}_{f}$ (it is assumed unobservable, so it must be moved
to negative infinity) to its location $\vec{x}$ along certain path
defined on a 2 dimensional space-like plane at given time $\tau$,\begin{equation}
\psi(\vec{x},\tau)=\lim_{\vec{x}_{f}\rightarrow-\infty}e^{-i\int_{\vec{x}_{f}}^{\vec{x}}dy^{i}a_{i}(\vec{y},\tau)},\label{eq:bosonize}\end{equation}
where $a_{i}$ is a vector field with subscript $i=1,2$. We can not
see the string itself, only the end point of the string, we can imagine
that a fermion always attaches an unobserved string, when the position
of the fermion changes, it brings along the attached string. The string
or the path in the formula is arbitrary, such phase arbitrariness
of $\psi$ reflects the fact that there are redundant degree of freedoms
for $\psi$ and $a_{i}$. This property indicates that the $a_{i}$
field is essentially a gauge field, one specific path gives a specific
gauge. The temporal component of the gauge field $a_{0}$, which is
not included in Eq.(\ref{eq:bosonize}), can be introduced by a gauge
condition. Then space-like plane mentioned above does not necessarily
be space-like, it can be any Lorentz rotated hyper-plane from the
spatial one in the 2+1 dimensions spacetime. And the starting fixed
point $\vec{x}_{f}$ at spatial negative infinity is now Lorentz rotated
to a new starting fixed point that is not necessarily space-like,
but a fixed point defined on the hyper-plane and then be moved to
the negative infinity of the hyper-plane. Thus now the gauge field
can be formally generalized to a covariant form, i.e. the subscript
now is $i=0,1,2$. We could write the fields in a convariant form
$\psi(x)$ and $a_{i}(x)$, $x=(\vec{x},\tau)$, and the gauge transformations
of them is given by\begin{equation}
\psi(x)\rightarrow e^{i\varphi(x)}\psi(x),\quad a_{i}(x)\rightarrow a_{i}(x)+\partial_{i}\varphi(x).\end{equation}
For massless fermions, there is another independent chiral fermionic
variable $\bar{\psi}$ constructed as\begin{equation}
\bar{\psi}(\vec{x},\tau)=\lim_{\vec{x}_{f}\rightarrow-\infty}e^{i\int_{\vec{x}_{f}}^{\vec{x}}dy^{i}\bar{a}_{i}(\vec{y},\tau)},\end{equation}
in which $\bar{a}_{i}$ denotes $\bar{a}(k)\equiv a(-k)$.

After performing the replacement of $\psi$ by the new field $a_{i}$,
then all operators appearing in the theory become bosonic, there will
be no fermions in our discussions any more. It is worth stressing
that all the paths in the definitions start from a fixed point $x_{f}$
(although it is unobserved at negative infinity) is very important
in reproducing the fermionic statistics. We will show the proof that
the fermionic anti-commutation relation of $\psi$ on equal time is
automatically ensured in Appendix I. Sometimes in our formulas we
do not write explicitly the starting point $x_{f}$ in the path integral
and then take the limit $x_{f}\rightarrow-\infty$, but just writing
the starting point of the path integral instead as $-\infty$, we
must keep in mind that all the paths integral from negative infinity
start from a negative infinite fixed point.

The bosonization recipe gives a dictionary translating the fermionic
terminologies to their bosonic counterparts as follows.

\begin{center}
\begin{tabular}{|c|c|c|}
\hline 
 & fermionic & bosonic\tabularnewline
\hline
\hline 
current $J_{i}$ & $\bar{\psi}\gamma_{i}\psi$ & $\pm\frac{1}{2\pi}i\epsilon_{ijk}\partial_{j}a_{k}$\tabularnewline
\hline 
kinetic energy & $i\bar{\psi}\gamma_{i}\partial_{i}\psi$ & $\pm\frac{1}{2\pi}i\epsilon_{ijk}a_{i}\partial_{j}a_{k}$\tabularnewline
\hline 
chemical potential $\mu J_{0}$ & $\mu\bar{\psi}\gamma_{0}\psi$ & $\pm\frac{1}{2\pi}i\mu\epsilon_{ij}\partial_{i}a_{j}$\tabularnewline
\hline 
source $A_{i}J_{i}$ & $A_{i}\bar{\psi}\gamma_{i}\psi$ & $\pm\frac{1}{2\pi}i\epsilon_{ijk}A_{i}\partial_{j}a_{k}$\tabularnewline
\hline 
chiral density $\mathcal{O}_{\pm}$ & $\bar{\psi}\left(1\pm\gamma_{5}\right)\psi$ & $M^{2}\exp\left(\pm i\Phi\right)$\tabularnewline
\hline 
mass term & $m\bar{\psi}\psi$ & $mM^{2}\cos\Phi$\tabularnewline
\hline
\end{tabular}
\par\end{center}

The bosonization recipe can be formally generalized to the non-Abelian
fermion, by replacing the Abelian gauge field $a_{i}$ by the non-Abelian
gauge field $a_{i}=a_{i}^{I}t^{I}$. Similarly, when the path defined
on a spatial hyper-plane, the bosonization is given by \begin{equation}
\psi(\vec{x},\tau)=\exp\left(-i\int_{-\infty}^{\vec{x}}dy^{i}t^{I}a_{i}^{I}(\vec{y},\tau)\right),\quad\bar{\psi}(\vec{x},\tau)=\exp\left(i\int_{-\infty}^{\vec{x}}dy^{i}t^{I}\bar{a}_{i}^{I}(\vec{y},\tau)\right)\end{equation}
where $t^{I}$ with $I=(1,2,...,N^{2}-1)$ are the generators of the
$SU(N)$ gauge group. The gauge transformations of the covariant form
of the fields are given by\begin{equation}
\psi(x)\rightarrow e^{it^{I}\varphi^{I}(x)}\psi(x),\quad a_{i}^{I}(x)\rightarrow a_{i}^{I}(x)+\partial_{i}\varphi^{I}(x)+\theta^{IJK}a_{i}^{J}\varphi^{K}(x),\end{equation}
in which $\theta^{IJK}$ is the structure constant of the gauge group
satisfying $[t^{I},t^{J}]=i\theta^{IJK}t^{K}$. Since there are $N^{2}-1$
gauge-invariant conserved currents, the number of chemical potentials
$\mu^{I}$ is generalized to $N^{2}-1$ correspondingly. The dictionary
for the non-Abelian fermions is as follows.

\begin{center}
\begin{tabular}{|c|c|c|}
\hline 
 & fermionic & bosonic\tabularnewline
\hline
\hline 
current $J_{i}^{I}$ & $\bar{\psi}\gamma_{i}t^{I}\psi$ & $\pm\frac{1}{2\pi}i\epsilon_{ijk}\left(\partial_{j}a_{k}^{I}+\theta^{IJK}a_{j}^{J}a_{k}^{K}\right)$\tabularnewline
\hline 
kinetic energy & $i\bar{\psi}\gamma_{i}\partial_{i}\psi$ & $\pm\frac{1}{2\pi}i\epsilon_{ijk}\left(a_{i}^{I}\partial_{j}a_{k}^{I}+\frac{2}{3}\theta^{IJK}a_{i}^{I}a_{j}^{J}a_{k}^{K}\right)$\tabularnewline
\hline 
chemical potential $\mu^{I}J_{0}^{I}$ & $\mu^{I}\bar{\psi}\gamma_{0}t^{I}\psi$ & $\pm\frac{1}{2\pi}i\mu^{I}\epsilon_{ij}\left(\partial_{i}a_{j}^{I}+\theta^{IJK}a_{i}^{J}a_{j}^{K}\right)$\tabularnewline
\hline 
source $A_{i}^{I}J_{i}^{I}$ & $A_{i}^{I}\bar{\psi}\gamma_{i}t^{I}\psi$ & $\pm\frac{1}{2\pi}i\epsilon_{ijk}A_{i}^{I}\left(\partial_{j}a_{k}^{I}+\theta^{IJK}a_{j}^{J}a_{k}^{K}\right)$\tabularnewline
\hline 
chiral density $\mathcal{O}_{\pm}$ & $\bar{\psi}\left(1\pm\gamma_{5}\right)\psi$ & $M^{2}\mathrm{tr}\exp\left(\pm i\Phi\right)$\tabularnewline
\hline 
mass term & $m\bar{\psi}\psi$ & $mM^{2}\mathrm{tr}\cos\Phi$\tabularnewline
\hline
\end{tabular}
\par\end{center}

The detail proofs of the two dictionaries will be given in the appendix-II
to -IV. We can see that the current is identically conserved, since
the current is just $J=\pm*f$ and $d*f=0$. The significant features
of the above translations are mathematically transparent. The first
fundamental feature is that the bosonized theory is very closely related
to Chern-Simons and/or topological field theory. The kinetic energy
term of fermions directly takes the Chern-Simons form, and the source
term which describes the interaction between fermion and gauge field
is of the mutual-Chern-Simons form. The term {}``mutual'' means
that the external gauge field $A_{i}$ and the bosonized gauge field
$a_{i}$ are mutually and topologically interacting with each other
in a Chern-Simons form.

We also observe in these dictionaries that the chiral density of fermions,
from the bosonic perspective, measures the flux density of the $a$
field, where the $\Phi(x)=\lim_{\Sigma_{x}\rightarrow0}\iint_{\Sigma_{x}}dx^{i}dx^{j}f_{ij}$
is the flux density (the non-Abelian case is just replacing $f_{ij}$
by $f_{ij}^{I}$). Here the energy scale $M^{2}=\frac{1}{2\pi\epsilon^{2}}$
is explicitly introduced to the system which sets the scale for the
chiral symmetry breaking. This energy gap is related to the Gribov
gap \cite{Gribov:1977wm} at low energy which inevitably exists in
quantum gauge theories especially in non-Abelian cases. As we will
see that this energy gap $M$ is responsible for a ordered chiral
symmetry breaking and protecting a low energy incompressible fluid-like
state at disorder, both of which are common low energy features in
a strongly coupled gauge system at finite chemical potential. This
reformulation of the chiral density (and the corresponding chiral
order parameter) provides a possibility to discuss the relation between
the chiral symmetry breaking and confined/deconfined phase transition
on an equal footing in section V.

For the practical interest, e.g. the QCD system and many strong correlated
electron systems in 2+1 dimensions, our main focus in this paper is
the massless fermions. By the construction, the action of a gauge
theory with massless fermion matters can be straightforwardly rewritten
as a pure bosonic action without any fermion fields,\begin{align}
Z & =\int\mathcal{D}A\mathcal{D}\bar{\psi}\mathcal{D}\psi\exp\left(-\int d^{3}x\bar{\psi}\left[i\gamma_{i}\left(\partial_{i}-iA_{i}\right)+\mu\gamma_{0}\right]\psi-S[A]\right)\label{eq:original_action}\\
 & =\int\mathcal{D}A\mathcal{D}a\left(\prod_{x}\bar{\psi}_{x}\psi_{x}\right)\exp\left(-S_{b}[a,A]\right),\nonumber \\
 & =M^{2}\int\mathcal{D}A\mathcal{D}a\left(\prod_{x}\cos\Phi_{x}\right)\exp\left(-S_{b}[a,A]\right)\nonumber \\
 & =M^{2}\int\mathcal{D}A\mathcal{D}a\exp\left(-S_{b}[a,A]+\sum_{x}\ln\cos\Phi_{x}\right)\nonumber \\
 & =M^{2}\int\mathcal{D}A\mathcal{D}a\exp\left(-S_{b}[a,A]+\sum_{x}\ln\left(1-\frac{1}{2}\Phi^{2}+...\right)\right)\nonumber \\
 & =M^{2}\int\mathcal{D}A\mathcal{D}a\exp\left(-S_{b}[a,A]-\frac{1}{8\pi^{2}M}\int d^{3}xf_{ij}f_{ij}+...\right)\nonumber \\
 & =M^{2}\int\mathcal{D}A\mathcal{D}a\exp\left(-S_{eff}[a,A]\right)\nonumber \end{align}
with\begin{align}
S_{eff} & =\pm\frac{1}{2\pi}i\int d^{3}x\left[\epsilon_{ijk}\left(a_{i}\partial_{j}a_{k}+A_{i}\partial_{j}a_{k}\right)+\mu\epsilon_{ij}\partial_{i}a_{j}\right]+S[A]+\frac{1}{8\pi^{2}M}\int d^{3}x\left(\partial_{i}a_{j}\right)^{2},\label{eq:abelian-action}\end{align}
in which $S[A]$ is the action of the original pure gauge fields $S[A]=-\frac{1}{4g^{2}}F_{ij}(A)F_{ij}(A)$,
$\mu$ is the chemical potential of the fermion, and the last term
comes from the change of the functional integral measure (i.e. $\prod_{x}\cos\Phi_{x}$)
expanding in powers of the small parameter $1/M$. The mass gap $M$
suppresses the Maxwell term of $a$ field, thus leading to a protection
of the topological behavior of the $a$ field due to the Chern-Simons
term. This gap originated from the requirement of a symmetry breaking
of the system, however, it is striking that such requirement directly
implies the existence of a topological order beyond the ordered symmetry
breaking phase, in which the gap $M$ will protect a fluid-like state
as its low energy modes described by the Chern-Simons term. Therefore,
the effective action Eq.(\ref{eq:abelian-action}) is at the vicinity
of the IR fixed point far below the scale $M$, thus describing the
low energy long wavelength behavior of the gauge systems.

The non-Abelian version of Eq.(\ref{eq:abelian-action}) is\begin{equation}
S_{eff}^{NA}=\pm\frac{1}{2\pi}i\mathrm{tr}\int d^{3}x\left[\epsilon_{ijk}\left(a_{i}^{I}f_{jk}^{I}+\frac{2}{3}\theta^{IJK}a_{i}^{I}a_{j}^{J}a_{k}^{K}+A_{i}^{I}f_{jk}^{I}\right)+\epsilon_{ij}\mu^{I}f_{ij}^{I}\right]+S[A]+\frac{1}{8\pi^{2}M}\mathrm{tr}\int d^{3}xf_{ij}^{I}f_{ij}^{I},\label{eq:nonabelian-action}\end{equation}
where $f_{ij}^{I}=\frac{1}{2}\left(\partial_{i}a_{j}^{I}+\theta^{IJK}a_{i}^{J}a_{j}^{K}\right)$
is the curvature of $a_{i}^{I}$. 

If we consider that fermions are massive, it is equivalent to introduce
the extra Lagrangian terms,\begin{align}
mM^{2}\mathrm{tr}\cos\Phi & =mM^{2}\mathrm{tr}\left(1-\frac{1}{2}\Phi^{2}+...\right)\nonumber \\
 & =mM^{2}-\frac{m}{8\pi^{2}M^{2}}\mathrm{tr}f_{ij}f_{ij}+...\end{align}
in which the first constant term is irrelevant, and the second term
is the conventional Maxwell/Yang-Mills term of $a_{i}$ fields.

The similar bosonized effective actions without chemical potential
are also found in literature, however, what we focus in the paper
is at their finite chemical potential regions. An important observation
to these effective actions is that the chemical potential just formally
shifts the temporal component of the gauge field at least in the Abelian
case \cite{Hasenfratz:1983ba}. It is obvious that since there is
no dynamics of the $a_{0}$ or $a_{0}^{I}$, it plays the role of
a Lagrange multiplier in the action which keeps the particle number
$J_{0}$ or $J_{0}^{I}$ conserved, similar with the chemical potential
$\mu$ or $\mu^{I}$. Therefore, we conclude that when chemical potential
is introduced in the gauge systems, the Chern-Simons term inevitably
emerges as a constraint, when fermions are bosonized. Only when the
fermions are reformulated by bosons this point can be easily seen.
This Chern-Simons term inevitably presenting at finite chemical potential
is protected by the finite gap $M$ and strong coupling of Yang-Mills
term, thus leading to a low energy topological dominant theory at
finite chemical potential, while the dynamical Maxwell/Yang-Mills
part plays the role of perturbation. 

One may caution that the Chern-Simons term although gauge invariant,
usually breaks the parity and time reversal symmetry. It is worth
emphasizing that there is an arbitrariness for prefactor $\pm i$
of the Chern-Simons action, which comes from the arbitrariness of
the phase in the anti-commutation relation ($ab-e^{\pm i\pi}ba=0$)
of the fermions (see the appendix I) and the bosonized form of the
current (see the appendix II). The arbitrariness makes the theory
does not break the parity and time reversal symmetry as it should,
when the Chern-Simons term changes sign under parity $\epsilon_{ijk}a_{i}\partial_{j}a_{k}\rightarrow-\epsilon_{ijk}a_{i}\partial_{j}a_{k}$,
and $it\rightarrow-it$ under time reversal. This property indicates
the particle/anti-particle symmetry ($\mu\leftrightarrow-\mu$) in
the phase diagram of the system. By convention, we only pick up the
plus sign in our following discussions. A salience change of the new
actions concerns the center symmetry $Z_{N}$ of the gauge symmetry
$SU(N)$. The bosonization makes the dynamic matter fields be adjoint
representations, so all the fields are $Z_{N}$-invariant, thus the
center symmetry does not break, which will discuss in Section IV.
To summarize, the bosonized effective action preserves all symmetries
of the origin action Eq.(\ref{eq:original_action}), and hence Eq.(\ref{eq:abelian-action})
and/or Eq.(\ref{eq:nonabelian-action}) serve a good starting point
for studying the gauge system at finite chemical potential in 2+1
dimensions.

\section{Sign Structure at Finite Chemical Potential}

In the quantum mechanical picture, the fermion sign problem arises
due to the fact that the many particle wavefunction changes sign when
any two fermions are interchanged. As a consequence, considering a
fermion in a fermionic density bath and travels around a closed spatial
trajectory $C$. In the path integral formalism, everytime a fermion
crossing (equivalent to exchange) an environment fermion gives an
extra minus sign to the statistic weight of the path integral, beside
the conventional part $Z_{0}[C]$ describing the zero density contribution.
So the partition function of the many fermions system is given by
$Z=\sum_{C}(-1)^{N[C]}Z_{0}[C]$, where $N[C]$ denotes the total
number of the fermions on the trajectory C that needs to cross. The
partition function sums over all possible closed loop C and takes
an extra sign structure. Compared with the zero density partition
function $Z=\sum_{C}Z_{0}[C]$, the statistic weight $(-1)^{N[C]}$
as a functional of the loop C in the summation is highly oscillatory.
This property makes it hard to evaluate the partition function numerically,
since the sign or the phase lead to dramatic cancellations which makes
statistical errors scale exponentially when the system approaches
to the thermodynamic limit.

In the field theoretical approach to the strongly coupled many particle
system, the fermion density is characterized by the value of the fermion
chemical potential or chemical potential $\mu$. When we integrate
out the fermions in the partition function, it gives a statistic weight
$\det D(\mu,A)$ to the path integral, where \begin{equation}
D(\mu,A)=i\gamma_{\mu}\left(\partial_{\mu}-iA_{\mu}\right)+\mu\gamma_{0}\end{equation}
is the Dirac operator. The notorious sign problem arises when the
fermion determinant is not positive but in general be a complex number
with non-trivial phase $\det D\equiv\left|\det D\right|e^{i\theta}$
at finite chemical potential. Several proposals have been trying to
at least partially solve the sign problem \cite{PhysRevLett.72.2442,PhysRevLett.83.3116,PhysRevB.71.155115,PhysRevLett.102.131601}.

The bosonized formulation provides us a new perspective to the fermion
sign problem in 2+1 dimensions. Take the Abelian case as an example.
In Eq.(\ref{eq:abelian-action}), we note that the action is quadratic
in $a_{i}$, so the bosonized matter fields $a_{i}$ can be integrated
out. It is equivalent to integrate out the matter fields, we obtain\begin{align}
Z & =\mathcal{N}\int\mathcal{D}A\exp\left(\int d^{3}x\left[\frac{i}{8\pi}\epsilon_{ijk}\left(A_{i}+\mu\delta_{0i}\right)\partial_{j}A_{k}-\frac{1}{32\pi^{2}M}F_{ij}(A)^{2}\right]-S[A]\right).\label{eq:abelian-integrate-out-fermion}\end{align}
One find that there is a new emergent Chern-Simons term for $A_{i}$
beside the original $S[A]$, which plays the role of the fermion determinant,
\begin{equation}
\det D(\mu,A)=\exp\left(-\frac{1}{32\pi^{2}M}\int d^{3}xF_{ij}(A)^{2}\right)\exp\left(\frac{i}{8\pi}\int d^{3}x\epsilon_{ijk}\left(A_{i}+\mu\delta_{0i}\right)\partial_{j}A_{k}\right).\end{equation}
Note that under homotopically non-trivial large gauge transformation
of $A$, the phase of the fermion determinant changes as $\det\left(\slash\!\!\!\partial+\slash\!\!\!\! A\right)\rightarrow\left(-1\right)^{n}\det\left(\slash\!\!\!\partial+\slash\!\!\!\! A\right)$
\cite{PhysRevLett.52.18,PhysRevD.29.2366} (n being an integer), which
can be absorbed into the sign arbitrariness in front of the Chern-Simons
action, and hence free from the anomaly. Therefore, the Chern-Simons
term consistently identifies with the phase of the fermion determinant.
While, the first exponent $\exp\left(-\frac{1}{32\pi^{2}M}\int d^{3}xF_{ij}^{2}\right)$
is real and it equals to the absolute value of the fermion determinant,
which renormalizes the gauge fields term $S[A]$, by $\frac{1}{g_{Re}^{2}}=\frac{1}{g^{2}}+\frac{1}{8\pi^{2}M}$.

In contrast to the original gauge fields term $S[A]$ which is metric
dependent (it is real in Euclidean spacetime while imaginary in Minkowski
spacetime), the Chern-Simons term is metric independent. This the
reason why Chern-Simons term is called topological. The consequence
of this topological property is that the Euclidean continuation does
not pick up any factor of $i$, since when we perform the Wick rotation
$t\rightarrow it$, the time component of the gauge field changes
as $A_{0}\rightarrow-iA_{0}$. Therefore, the action always has an
intrinsic imaginary part, no matter in Euclidean or Minkowski spacetime.
We come to a crucial conclusion that the non-trivial phase capturing
the essential of the sign problem is topological origin in 2+1 dimensions.
This connection between the notion of the zero temperature finite
chemical potential physics and the topology gives a further support
to the idea of topological order/phase that beyond the Landau's theories
of finite temperature phase transition. 

Only for special cases the theory could be free from the sign problem.
When we choose the temporal-axial gauge ($A_{0}=0$) which is trivial
for the reason that it is equivalent to a pure gauge system without
chemical potential. Or when $\mu$ is very large (compared with $A_{0}$)
to be dominant, together with performing an analytic continuation
of $\mu$ from real to imaginary, since now $\frac{1}{8\pi}\left|\mu\right|\epsilon_{ij}\partial_{i}A_{j}$
is positive \cite{de2002qcd,PhysRevD.67.014505,Lombardo:2006yc}.
However, in general, this method has not led to any improvement for
the sign problem, the non-trivial phase coming from fermionic statistic
is still there, we just transform the non-trivial phase into a generally
under-controlled topological gauge fields structure through bosonization.
The non-trivial phase in the statistic weight of the path integral
now has a complete new physical meaning, connecting to Chern-Simons
term, so we could manipulate the integral according to the properties
of the Chern-Simons theory, the sign problem in this sense is alleviated.
Let us consider an index usually measuring the badness of the sign
problem \begin{equation}
\left\langle S\right\rangle =\left\langle e^{i\theta}\right\rangle =Z^{-1}\int\mathcal{D}Ae^{iS_{CS}}e^{-S[A]}=\left\langle e^{iS_{CS}}\right\rangle ,\end{equation}
where $e^{i\theta}$ is defined as $\det D\equiv\left|\det D\right|e^{i\theta}$.
If it is much less than one then the sign problem is thought severe.
At high temperature, the system is $S[A]$ dominant, so the sign problem
is not severe, since $\left\langle S\right\rangle \sim Z^{-1}\int\mathcal{D}Ae^{-S}\approx1$.
And since the Chern-Simons action is an infrared fixed point action
and becomes dominant only at low temperature, the expectation value
gives $\left\langle S\right\rangle \sim Z^{-1}\int\mathcal{D}Ae^{iS_{CS}}$,
which is a topological invariant only depending on the topological
of the 3-manifold, not on the detail spatial size or temperature (temporal
size) of the system. In particular, for certain 3-manifold this quantity
may not be small, e.g. $\mathcal{M}=S^{2}\times S^{1}$, $\left\langle S\right\rangle \sim Z^{-1}\int\mathcal{D}Ae^{iS_{CS}}\approx1$
for any gauge group. The sign problem related to the Chern-Simons
term is irrelevant to the thermodynamics property we are interested
in. 

On the other hand, the low energy behavior governed by the Chern-Simons
term is protected by an energy gap proportional to the square of the
gauge coupling and $M$, so it is well protected especially at strong
coupling. Therefore, we can first safely integrate out the infrared
part which eliminates most of the sign problem, and have a factorization
of the integral into a topological (size and temperature independent)
part and a dynamic (size and temperature dependent, free from sign
problem) part, \begin{align}
\left\langle \mathcal{O}\right\rangle  & =Z^{-1}\int\mathcal{D}A\mathcal{O}(dA)e^{iS_{CS}[A]}e^{-S[dA]}\nonumber \\
 & =Z^{-1}\int\mathcal{D}\left(A_{IR}+B\right)\mathcal{O}(dA_{IR}+dB)e^{iS_{CS}[A_{IR}+B]}e^{-S[dA_{IR}+dB]}\nonumber \\
 & \approx Z^{-1}e^{iS_{CS}[A_{IR}]}\int_{\left|k\right|<\Lambda}\mathcal{D}B(k)e^{iS_{CS}[B]}\int_{\left|k\right|>\Lambda}\mathcal{D}B(k)\mathcal{O}(dB)e^{-S[dB]}\nonumber \\
 & =\mathrm{topological\; invariant}\times\int\mathcal{D}A\mathcal{O}(dA)e^{-S[dA]}\label{eq:factorization}\end{align}
where $\Lambda=g^{2}/\pi$ is the energy gap of $A$, $A_{IR}$ are
a set of flat gauge connections ($dA_{IR}=0$) dominating the infrared
behaviors. $\mathcal{O}(dA)$ is a gauge invariant thermodynamic observable
which is thought temperature and/or other thermodynamic variables
dependent and low energy topology irrelevant, so we formally put it
into the dynamic part of the integral. In principle, if the observable
can also be factorized according to the energy gap, it could contribute
to the low energy part of the integral, e.g. the Wilson loop, which
we will discuss in the next Section.

The factorization is based on the well separated scales of the topological
part and the high energy dynamic part. The stronger the gauge coupling,
the larger the gap between these two part, the more exact the factorization.
This factorization gives an approximation to hive off the Chern-Simons
term, which is relevant to the sign problem but irrelevant to the
thermodynamic property. If we are interested in the thermodynamical
behaviors of $\left\langle \mathcal{O}\right\rangle $, this approximation
may be useful.

\section{Wilson's Confinement Criteria at Finite Chemical Potential}

The most interesting physics happens in the finite chemical potential
region of gauge systems. It is known that in 2+1 dimensions the pure
gauge field theories only have one phase, i.e. confinement \cite{Polyakov:1976fu,Polyakov:111432}.
The deconfined transition occurs in the finite chemical potential
regime. However, a theoretical problem is, if the fermion matter fields
$\psi$ in the fundamental representations are included, the system
will lose its center invariance. Only when all the dynamical fields
are center invariant, the Wilson loop serves as an order parameter
to distinguish the confined and deconfined phases. As a consequence
the Wilson loop can not be used anymore to characterize these two
phases in this situation. On the other hand, the sign problem makes
the numerical method accessing the finite chemical potential region
very difficult. So we confront a difficulty how to study the confined
and deconfined phase at finite chemical potential.

The bosonization recipe is a promising approach, at least partially
alleviating the sign problem, to investigate the physics in the finite
chemical potential region. First of all, it bridges a gauge field
theory at finite chemical potential to a pure gauge field theory,
and hence, numbers of existing beautiful results and well developed
techniques for pure gauge system could be straightforwardly generalized
to finite chemical potential. Second, in contrast to the fact that
the matter fields $\psi$ in fundamental representation are not $Z_{N}$
invariant, the bosonized matter fields $a$ in adjoint representations
are $Z_{N}$ invariant. Therefore, all fields in the effective theory
are $Z_{N}$ invariant. This property lays the foundation to the validity
of Wilson loop being an order parameter, even when the dynamical matters
are included. This framework brings us a proper approach to discuss
the issue of confined/deconfined transition at finite chemical potential
which the fermionic framework can not do. Besides that, the Wilson
loop is a suitable observable for applying the factorization Eq.(\ref{eq:factorization}),
since the operator itself is factorisable $W(A_{IR}+B)=W(A_{IR})W(B)$,
the low energy sector of integration just gives the thermodynamic
independent knot polynomial \cite{Witten:1988hf}.

Let us briefly recall the results from the pure gauge systems. It
is well known that the expectation value of the Wilson loop satisfies
the area law at strong gauge coupling limit \cite{wilson1974confinement},
\begin{equation}
\left\langle W[C]\right\rangle =\int\mathcal{D}A\exp\left(i\oint_{C}A\right)\exp\left(iS[A]\right)\sim\exp\left(-\frac{1}{2}\sigma\mathcal{A}\right),\end{equation}
where $\sigma$ is the string tension at zero chemical potential,
and $\mathcal{A}$ is the minimal area enclosed by a loop $C$. The
area law of the Wilson loop implies a confining linear potential in
the static limit $L_{t}\rightarrow\infty$, i.e. $V=-\lim_{L_{t}\rightarrow\infty}\frac{1}{L_{t}}\ln\left\langle W\right\rangle =\frac{1}{2}\sigma r$,
in which we have set $\mathcal{A}=rL_{t}$, where $r$ and $L_{t}$
are the spatial and temporal lengths of the loop.

At finite chemical potential, the correction to such well known behavior
of Wilson loop could be considered as follows. Starting from the action
Eq.(\ref{eq:abelian-integrate-out-fermion}), we note that the Abelian
action $S[A]$ is $A_{i}$ shift invariant, so the only physical effect
of the chemical potential in Abelian gauge theories is just a shift
of the time component of the gauge field $A_{0}$. The Eq.(\ref{eq:abelian-integrate-out-fermion})
becomes\begin{equation}
Z=\int\mathcal{D}A\exp\left(\frac{i}{8\pi}\int d^{3}x\epsilon_{ijk}\tilde{A}_{i}\partial_{j}\tilde{A}_{k}-S[\tilde{A}]\right),\label{eq:action-no-a}\end{equation}
where $\tilde{A}_{i}=A_{i}+\mu\delta_{0i}$. At strong coupling limit,
the infrared behavior governed by the Chern-Simons term is well protected
by a gap. In this case, the Chern-Simons term which is responsible
for the sign problem can be first safely integrated out which gives
a topological invariant (knot polynomial). Thanks to its topological
nature, the topological invariant does not depend on the length or
size of the loop, the contribution from the Chern-Simons term does
not connect to the confining potential we are interested in. We can
ignore the Chern-Simons term in Eq.(\ref{eq:action-no-a}) and perform
the calculation of the Wilson loop as follows (in Minkowski background
with real time $t\in[0,L_{t}]$),\begin{equation}
\left\langle W[C,\mu]\right\rangle =\int\mathcal{D}A\exp\left(i\oint_{C}A\right)\exp\left(iS[\widehat{A}]\right),\end{equation}
where $\widehat{A}_{i}=A_{i}-i\mu\delta_{0i}$. Because the path integral
measure $\mathcal{D}A$ is shift invariant, we have\begin{align}
\left\langle W[C,\mu]\right\rangle  & =\int\mathcal{D}\widehat{A}\exp\left(i\oint_{C}A\right)\exp\left(iS[\widehat{A}]\right).\end{align}
By performing a variable replacement $\widehat{A}\rightarrow A$,
then $A_{i}$ becomes $A_{i}+i\mu\delta_{0i}$, we obtain\begin{align}
\left\langle W[C,\mu]\right\rangle  & =\int\mathcal{D}A\exp\left(i\oint_{C}dx^{j}\left(A_{j}+i\mu\delta_{0j}\right)\right)\exp\left(iS[A]\right)\nonumber \\
 & =\exp\left(i\oint_{C}dx^{j}\left\langle A_{j}+i\mu\delta_{0j}\right\rangle -\frac{1}{2}\oint_{C}dx^{i}\oint_{C}dy^{j}\widehat{K}_{ij}(x-y)\right),\end{align}
For $\left\langle A_{j}\right\rangle =0$, the first part in the exponent
gives \begin{equation}
i\oint_{C}dx^{j}i\mu\delta_{0j}=-\mu L_{t}.\label{eq:Wilson-1}\end{equation}
The $\widehat{K}_{ij}(x-y)\equiv\left\langle T\left(A_{i}(x)+i\mu\delta_{0i}\right)\left(A_{j}(y)+i\mu\delta_{0j}\right)\right\rangle $
is the correlator of the shift gauge field, which is given by\begin{align}
\widehat{K}_{ij}[\widehat{A}_{i}] & =\widehat{K}_{ij}[A_{i}+i\mu\delta_{0i}]\nonumber \\
 & =K_{ij}[A_{i}]+i\mu\frac{\delta K_{ij}}{\delta A_{0}}-\mu^{2}\delta_{0k}\delta_{0l}\frac{\delta^{2}K_{ij}}{\delta A_{k}\delta A_{l}}\nonumber \\
 & =K_{ij}(x-y)-\mu^{2}P_{ij}^{L}(x-y),\end{align}
where $P_{ij}^{L}$ is the longitudinal projector. The unshifted part
correlator $K_{ij}$ gives rise to the standard area law at strong
coupling limit, \begin{equation}
-\frac{1}{2}\oint_{C}dx^{i}\oint_{C}dy^{j}K_{ij}=-\frac{1}{2}\sigma\mathcal{A}=-\frac{1}{2}\sigma rL_{t}.\label{eq:Wilosn-2}\end{equation}
The shift part $-\mu^{2}P_{ij}^{L}$ gives a correction to the existing
area law behavior. By using the Stokes's theorem, we get\begin{align}
-\frac{1}{2}\oint_{C}dx^{i}\oint_{C}dy^{j}\left(-\mu^{2}P_{ij}^{L}\right) & =\frac{1}{2}\mu^{2}\int_{\Sigma}d^{2}x\int_{\Sigma}d^{2}y\epsilon_{ik}\epsilon_{jl}\partial_{k}\partial_{l}P_{ij}^{L}\nonumber \\
 & =\frac{1}{2}\mu^{2}\int_{\Sigma}d^{2}x\int_{\Sigma}d^{2}y\left(\delta_{ij}\partial^{2}-\partial_{i}\partial_{j}\right)P_{ij}^{L},\end{align}
where $\Sigma$ is the minimal surface enclosed by the loop C. According
to the Ward identity $\partial_{i}P_{ij}^{L}=0$, and the Possion
equation $\partial^{2}P_{ij}^{L}(x)=\delta_{ij}\delta^{2}(x)$ in
2+1 dimensions, we obtain\begin{align}
\frac{1}{2}\mu^{2}\int_{\Sigma}d^{2}x\int_{\Sigma}d^{2}y\left(\delta_{ij}\partial^{2}-\partial_{i}\partial_{j}\right)P_{ij}^{L} & =\frac{1}{2}\mu^{2}\int_{\Sigma}d^{2}x\int_{\Sigma}d^{2}y\delta^{2}(x-y)\nonumber \\
 & =\frac{1}{2}\mu^{2}\mathcal{A}=\frac{1}{2}\mu^{2}rL_{t}.\label{eq:Wilson-3}\end{align}
Put all Eq.(\ref{eq:Wilson-1},\ref{eq:Wilosn-2},\ref{eq:Wilson-3})
together, we finally have\begin{equation}
\left\langle W[C,\mu]\right\rangle \sim\exp\left[-\mu L_{t}-\frac{1}{2}\left(\sigma-\mu^{2}\right)\mathcal{A}\right].\label{eq:area-law-abelian}\end{equation}
where {}``$\sim$'' means, at strong gauge coupling limit, the result
asymptotically behaves as the following form. The static confining
potential at finite chemical potential is therefore obtained\begin{equation}
V=-\lim_{L_{t}\rightarrow\infty}\frac{1}{L_{t}}\ln\left\langle W\right\rangle =\mu+\frac{1}{2}\left(\sigma-\mu^{2}\right)r.\label{eq:potential-abelian}\end{equation}
The first term of the static potential naturally comes from the chemical
potential, the second term is a confining linear potential with a
chemical potential dependent string tension $\sigma\rightarrow\sigma-\mu^{2}$.
It is very physically importance for the minus sign, since it makes
the chemical potential play the role of a competition against the
confinement, i.e. deconfinement. The chemical potential weakens the
string tension, and finally snaps the string at a critical value $\mu_{c}=\sigma^{1/2}$. 

At finite temperature, the string tension is in general a function
of temperature. Consider the dimension of the string tension is of
energy squared, so near the critical temperature we have the behavior,\begin{equation}
\sigma(T)\sim\sigma_{0}T_{c}^{2}\left(1-\frac{T^{\eta}}{T_{c}^{2}}\right),\end{equation}
in which $T_{c}$ is the critical temperature at zero chemical potential
($\mu=0$), the $\eta$ is the critical exponent near $T_{c}$, and
$\sigma_{0}$ is a constant. Then the chemical potential and temperature
dependence of the string tension is given by\begin{equation}
\sigma(T,\mu)=\sigma_{0}T_{c}^{2}\left(1-\frac{T^{\eta}}{T_{c}^{2}}-\frac{\mu^{2}}{\sigma_{0}T_{c}^{2}}\right)=\sigma_{0}\widehat{T}_{c}^{2}\left(1-\frac{T^{\eta}}{\widehat{T}_{c}^{2}}\right),\end{equation}
where the chemical potential dependence of the critical temperature
is found \begin{equation}
\widehat{T}_{c}(\mu)=T_{c}\left(1-\frac{\mu^{2}}{\sigma_{0}T_{c}^{2}}\right)^{\frac{1}{\eta}}.\label{eq:phase-boundary-abelian}\end{equation}
This relation depicts the confined/deconfined phase transition boundary
in the $\mu-T$ phase diagram. The critical chemical potential is
predicted at $\mu_{c}=\sigma_{0}^{1/2}T_{c}$, i.e. ratio between
the critical chemical potential and the critical temperature in confined/deconfined
transition is a constant.

The above result could be generalized to the non-Abelian case, which
is not very trivial. Compared with the Abelian case, the similar aspects
comes from the fact that the chemical potential term $\mu^{I}J_{0}^{I}$
is also able to absorbed into the source term $A_{i}^{I}J_{i}^{I}$,
and so we could treat the chemical potential as a shift to the time
component of the non-Abelian gauge field, $A_{i}^{I}\rightarrow A_{i}^{I}+\mu^{I}\delta_{0i}$.
However, the physical effects of chemical potential would no longer
be just the formally shift of $A_{0}^{I}$, like the Abelian case.
In other words, the shift will introduces extra terms due to its intrinsic
non-linearity, which do not appear in the Abelian gauge theories,\begin{align}
S_{YM}[A]+A_{i}^{I}J_{i}^{I}+\mu^{I}J_{0}^{I} & =S_{YM}[A]+\tilde{A}_{i}^{I}J_{i}^{I}\neq S_{YM}[\tilde{A}]+\tilde{A}_{i}^{I}J_{i}^{I}\nonumber \\
 & =S_{YM}[\tilde{A}]+\tilde{A}_{i}^{I}J_{i}^{I}+\mathcal{O}(\mu^{4})+\mathcal{O}(A\mu^{3})+\mathcal{O}(A^{2}\mu^{2})+\mathcal{O}(A^{3}\mu).\label{eq:extra-terms}\end{align}
 The term $\mathcal{O}(\mu^{4})$ is a constant, and the terms $\mathcal{O}(A\mu^{3})$
do not give correction to the correlator of $A_{i}^{I}$, so both
are irrelevant to the physics. The terms $\mathcal{O}(A^{2}\mu^{2})$
are vanished, i.e.\begin{align}
\theta^{IJK}A_{i}^{J}\mu^{K}\delta_{0j}\theta^{ILM}A_{i}^{L}\mu^{M}\delta_{0j}+\theta^{IJK}A_{i}^{J}A_{j}^{K}\theta^{ILM}\mu^{L}\mu^{M}\delta_{0i}\delta_{0j} & \equiv0\end{align}
Only $\mathcal{O}(A^{3}\mu)$ terms are non-vanished. But note that
this term always has $A_{0}^{I}$ due to the contraction between $A_{i}^{I}$
and $\mu^{J}\delta_{0i}$, therefore, if we choose the temporal-axial
gauge to eliminate $A_{0}$, this term vanishes as well. To summarize,
under the temporal-axial gauge, these extra terms in Eq.(\ref{eq:extra-terms})
are irrelevant to what we are interested in. So the reasoning in the
Abelian case still holds, just replacing the $K_{ij}$ by $K_{ij}^{IJ}\equiv\left\langle TA_{i}^{I}A_{j}^{J}\right\rangle $
which takes the value under the temporal-axial gauge. As we know that,
because the Wilson loop is gauge invariant, so the temporal-axial
gauge correlator will also safely give the reliable result. 

Thus, similar with what we have done in the Abelian case, we ignore
the (non-Abelian) Chern-Simons term which gives rise to loop-size-irrelevant
topological invariant, and the Wilson loop can be calculated as,\begin{align}
\left\langle W[C,\mu]\right\rangle  & =\int\mathcal{D}A\left[\frac{1}{N}\mathrm{tr}\exp\left(i\oint_{C}dx^{j}A_{j}^{I}t^{I}\right)\right]\exp\left(iS[\widehat{A}]\right)\nonumber \\
 & =\int\mathcal{D}A\left[\frac{1}{N}\mathrm{tr}\exp\left(i\oint_{C}dx^{j}\left(A_{j}^{I}+i\mu^{I}\delta_{0j}\right)t^{I}\right)\right]\exp\left(iS[A]\right)\nonumber \\
 & =\exp\left(\frac{i}{N}\mathrm{tr}\oint_{C}dx^{j}\left\langle A_{j}^{I}+i\mu^{I}\delta_{0j}\right\rangle t^{I}-\frac{1}{2N}\mathrm{tr}\oint_{C}dx^{i}\oint_{C}dy^{j}\widehat{K}_{ij}^{IJ}t^{I}t^{J}\right).\end{align}

Unlike the Abelian case, the first term in the exponent is vanished
for the tracelessness of the group generators, \begin{equation}
\frac{i}{N}\mathrm{tr}\oint_{C}dx^{j}\left\langle A_{j}^{I}+i\mu^{I}\delta_{0j}\right\rangle t^{I}=\frac{i}{N}\mathrm{tr}\left(t^{I}\right)\oint_{C}dx^{j}i\mu^{I}\delta_{0j}=0.\end{equation}
This result exhibits the consequence that if there are several chemical
potential components in the non-Abelian gauge theories, they will
cancel each other in the density bath and do not contribute to the
averaged static potential.

The second term in the exponent is given by\begin{align}
-\frac{1}{2N}\mathrm{tr}\oint_{C}dx^{i}\oint_{C}dy^{j}\widehat{K}_{ij}^{IJ} & =-\frac{1}{2N}\mathrm{tr}\left(t^{I}t^{J}\right)\oint_{C}dx^{i}\oint_{C}dy^{j}\left(K_{ij}^{IJ}-\mu^{I}\mu^{J}P_{ij}^{L}\right)\nonumber \\
 & =-\frac{1}{2}\left(\sigma-\frac{1}{N}\sum_{I}^{N^{2}-1}\mu_{I}^{2}\right)\mathcal{A}.\end{align}
Put them together and we find the non-Abelian generalization of Eq.(\ref{eq:area-law-abelian})\begin{equation}
\left\langle W[C,\mu]\right\rangle \sim\exp\left[-\frac{1}{2}\left(\sigma-\frac{1}{N}\sum_{I}^{N^{2}-1}\mu_{I}^{2}\right)\mathcal{A}\right].\label{eq:area-law-nonabelian}\end{equation}
It straightforwardly gives the static potential\begin{equation}
V=\frac{1}{2}\left(\sigma-\frac{1}{N}\sum_{I}^{N^{2}-1}\mu_{I}^{2}\right)r,\end{equation}
and the chemical potential dependence of the critical temperature,
i.e. non-Abelian version of Eq.(\ref{eq:phase-boundary-abelian})
is given by \begin{equation}
\widehat{T}_{c}(\mu)=T_{c}\left(1-\frac{\frac{1}{N}\sum_{I}^{N^{2}-1}\mu_{I}^{2}}{\sigma_{0}T_{c}^{2}}\right)^{\frac{1}{\eta}}.\label{eq:phase-boundary-nonabelian}\end{equation}

To summarize, up to a topological invariant, we calculate the chemical
potential dependence of the Wilson loop at strong coupling limit for
the Abelian and non-Abelian gauge theories in 2+1 dimensions. It should
be emphasized two key points here. First, the chemical potential formally
shifts the time-component gauge field. In the Abelian case, that is
all, but in the non-Abelian case, beside the shift it introduces several
extra terms due to its intrinsic non-linearity. However, these extra
terms coming from the chemical potential shift are proportional to
the {}``dynamic part'' of $A_{0}$-component gauge field, so they
could be gauged away and do not contribute to our final observables.
Second, the chemical potential formally shifts the area law, or equivalently
the string tension with a minus sign. The area law induced by the
chemical potential comes from the fact that the chemical potential
gives a non-trivial longitudinal part to the correlator which could
not be gauged away and has crucial physical effect. It is intuitive
reasonable that the gauge field gains its non-trivial longitudinal
component in a density bath. The confinement or the area law is looked
like that the gauge fields are gaped in a specific way so that they
could no longer be excited, only the longitudinal part is responsible
for the gap (note the massive photon propagator behaves like a longitudinal
projector in the large momentum limit) and breaks the gauge symmetry
in the ground-state (spontaneously). The minus sign signifies that
the effect of the density bath is anti-gaped at strong coupling, which
indicates the deconfinement. The anti-gaped effect of the density
bath at strong-coupling may play a similar role of an instability
that possibly develops (e.g. superconducting) phase transition near
the confined/deconfined phase boundary.

\section{Relation Between Chiral Symmetry Breaking and Confinement}

Besides the under-controlled sign problem, the generalization of Wilson's
confinement criteria at finite chemical potential mentioned in previous
sections, we will refer to another advantage of the bosonization recipe
in this section, this recipe provides us a suitable method to discuss
the one of the long-standing puzzles, the relation between the chiral
symmetry breaking and confinement. 

As is discussed in the introduction, the chiral symmetry breaking
phase is an old story through the Landau-Ginzburg's framework, in
which the phase is completely characterized by the local order parameter
and classified by the breaking chain of symmetry group, and we have
Goldstone bosons as the low energy excitations in the phase. The fermionic
language (e.g. BCS-like NJL model) is good at such type of description
and be our standard understanding of spontaneous chiral phase transition.
However, the gauge (gluonic) degrees of freedom are integrated out
and replaced by a local four-point interaction of fermions (quarks)
currents. As a consequence, the local $SU(N)$ gauge invariance is
replaced by a global $SU(N)$ symmetry in this type of models, so
that the property of confinement is lost, this fermionic type of models
are fail in describing the confined/deconfined phase transition. On
the other hand, the confined/deconfined phases are thought beyond
the Landau-Ginzburg's framework, since, strictly speaking, there seems
no local order parameter or symmetry breaking to distinguish these
two phases, especially at finite chemical potential. In order to investigate
the confined/deconfined phases deeply, the bosonic gauge theories
pictures are inevitable. Thanks to the advantage of the bosonization,
it helps us at least partially overcome the shortcomings at finite
chemical potential and defining a local chiral order parameter, both
of which are straightforward in the fermionic picture. Therefore,
we propose that the bosonization recipe is a proper approach being
able to treat the chiral symmetry breaking and the confinement on
an equal footing.

From the dictionary, one can see that the chiral density of fermions
in 2+1 dimensions, from the bosonic language, is just the flux vortex
density of bosonized $a$ fields. So the chiral symmetry breaking
phase is characterized by the non-vanishing (local) chiral order parameter\begin{equation}
\left\langle \bar{\psi}(x)\psi(x)\right\rangle =M^{2}\left\langle \mathrm{tr}\mathrm{\cos\left(\Phi(x)\right)}\right\rangle \equiv M^{2}\left\langle \mathrm{tr}U(x)\right\rangle \neq0.\end{equation}

On the other hand, because the system Eq.(\ref{eq:abelian-action})
and/or Eq.(\ref{eq:nonabelian-action}) is a pure gauge system with
two types of gauge fields $a$ and $A$, thus the system is confined
only when both pure gauge fields are confined. By using the Wilson's
criteria of confinement, we have,\begin{align}
\left\langle W[C]\right\rangle  & \approx\left\langle W[A,C]\right\rangle \left\langle W[a,C]\right\rangle \sim\exp\left(-\frac{1}{2}\sigma_{A}\mathcal{A}_{C}\right)\exp\left(-\frac{1}{2}\sigma_{a}\mathcal{A}_{C}\right)\label{eq:a-A-area-law}\end{align}
The relation qualitatively reproduces the behaviors of Eq.(\ref{eq:area-law-abelian})
and/or Eq.(\ref{eq:area-law-nonabelian}). The above two points are
standard results we have known. In this section, one of the goals
is to study the interplay between these two points.

In the flux vortex picture, suppose that we can subdivide any surface
enclosing by a loop into smaller surfaces, such that if a loop $C$
contains smaller loops $C_{1}$ and $C_{2}$, $\left\langle W[C]\right\rangle \approx\left\langle W[C_{1}]\right\rangle \left\langle W[C_{2}]\right\rangle $.
The less correlatedness between the loops the more exact the approximation.
Then consider a surface of planar loop C with area $\mathcal{A}$.
The surface C encloses a number ($\mathcal{A}/\mathcal{A}_{min}$)
of surfaces that each is bounded by $C_{i}$ with unit area $\mathcal{A}_{min}=\epsilon^{2}$.
The flux vortex $\left\langle \mathrm{tr}U_{i}(x)\right\rangle $
pierces the surface at point $x$ within loop $C_{i}$. For the Stokes's
theorem, then the Wilson loop C area pierced by these flux vortices
behaves like\begin{equation}
\left\langle W[a,C]\right\rangle =\left\langle \prod_{i}^{\mathcal{A}/\mathcal{A}_{min}}\mathrm{tr}U_{i}\right\rangle \approx\prod_{i}^{\mathcal{A}/\mathcal{A}_{min}}\left\langle \mathrm{tr}U_{i}\right\rangle =\left\langle \mathrm{tr}U\right\rangle ^{\frac{\mathcal{A}}{\mathcal{A}_{min}}}.\label{eq:product-vortices}\end{equation}
From this formula, we can see clearly that only when $\left\langle \mathrm{tr}U\right\rangle \neq0$
the behavior of the Wilson loop possesses an area law. It indicates
an important conclusion of the paper that the chiral symmetry breaking
$\left\langle \mathrm{tr}U\right\rangle =M^{-2}\left\langle \bar{\psi}(x)\psi(x)\right\rangle \neq0$
is a necessary condition for the confinement in 2+1 dimensions, i.e.
confinement implies chiral symmetry breaking. In a phase diagram for
the 2+1 dimensional gauge systems at finite chemical potential, this
result leads to the consequence that a confined phase is always contained
inside a chiral symmetry breaking phase regime. There could also exists
a deconfined phase in the chiral symmetry breaking phase, in which
fermion pairs have been formed, a gap coming from the pairing is opened
up, but the system is deconfined. The deconfinement phase could coexist
with the chiral symmetry breaking phase. It qualitatively behaves
like the pseudo-gap phenomenon in 2+1 dimensional doped cuprates \cite{1999RPPh...62...61T,2008RPPh...71a2501L}.
It is worthwhile to mention that this property is in contrast to the
counterpart in the 3+1 dimensional QCD deduced from large-N, in which
the authors argued that there is a phase in QCD system the confined
phase coexisting with the chiral restored phase, called quarkyonic
phase \cite{McLerran:2007qj}.

In our argument, there could exist both confined and deconfined phases
in the chiral symmetric broken phase, so let us discuss these two
phases based on the non-vanishing chiral order parameter/flux vortices
density. For the flux $\Phi$ is real, so $0<\left\langle \mathrm{tr}U\right\rangle <1$,
the string tension in Eq.(\ref{eq:a-A-area-law}) \begin{equation}
\sigma_{a}=-\frac{2\ln\left\langle \mathrm{tr}U\right\rangle }{\mathcal{A}_{min}}=-4\pi M^{2}\ln\left\langle \mathrm{tr}U\right\rangle \end{equation}
is positive defined, which means that the Wilson loop contributed
from the matter part possesses an area law decay and the matter field
is confined. Although the chiral condensation breaks gauge symmetry,
under a large gauge transformation, the system has a residue center
symmetry for the chiral order parameter $\left\langle \mathrm{tr}\cos\Phi\right\rangle \rightarrow\left\langle \mathrm{tr}\cos\left(\Phi+\frac{2\pi}{N}\cdot n\right)\right\rangle $,
with $n$=integer. The gauge symmetry is broken down to its discrete
center group, i.e. $SU(N_{f})_{L}\times SU(N_{f})_{R}\rightarrow SU(N_{f})\rightarrow Z_{N}$.
Therefore, we conclude that, although the 2+1 dimensional gauge theory
alone is in a confined phase, the system can also be confined when
matter is included, it is possible only when the matter fields are
condensed in such a way that the gauge symmetry is spontaneously broken
down to its discrete subgroup. 

The system also has a deconfined phase in another situation. The flux
vortices piercing the loop surface could wind in (Euclidean) time
direction by virtue of the periodic boundary condition. This phenomenon
frequently happens when the radius of the winding flux vortex loop
is smaller than that of the Wilson loop C, so that they will usually
pierce the loop surface twice in opposite directions inside the Wilson
loop, thus leading to a cancellation of these two contributions. Only
those flux vortex loops near the boundary of the Wilson loop surface
could pierce it once, which gives a perimeter law, \begin{equation}
\left\langle W[a,C]\right\rangle \sim\exp\left(-\alpha L_{C}\right),\end{equation}
where $L_{C}$ is the perimeter of the Wilson loop C, $\alpha$ is
certain constant. The behavior of perimeter law signifies a deconfinement.
Note that the center symmetry is not locally broken due to the non-vanished
local center flux vortices $\left\langle \mathrm{tr}U_{i}\right\rangle $.
The deconfined transition comes from the change of the global properties
of the center vortices in a finite size regime, in particular, their
size, shape and winding in the Wilson loop. In this sense, the confinement
and deconfinement are not characterized by their local symmetries,
but distinguished by the their global topological properties, the
confined/deconfined phase transition is beyond the Landau-Ginzburg
type of phase transition.

Generally speaking, there are two ways {}``seeing'' the flux vortices
in the system. One way is seeing the local vacuum expectation value
of the flux vortices, which characterizes the chiral symmetry breaking
phase in our bosonization picture. Another way is placing a finite
size Wilson loop in the system, and seeing the behavior of the Wilson
loop enclosing the flux vortices. The non-local behavior of it detect
the size, shape and winding properties of these flux vortices contained
inside the loop, which characterizing the confinement and deconfinement
phase. 

To summarize this section. In 2+1 dimensions, the chiral symmetry
breaking is a necessary condition for the confinement, but not sufficient.
The confined phase is always contained inside the chiral symmetry
breaking phase, and the deconfined phase could also coexist with the
chiral symmetry breaking phase. The non-vanishing center vortices
takes value of the center group. When the center vortices are long,
they pierce the Wilson loop surface only once, it gives an area law
and be confined. When the center vortices are short and winding to
form small loops, they usually pierce the Wilson loop surface twice
in opposite directions and cancel each other, it gives an perimeter
law and indicates a deconfined phase. The confined/deconfined transition
does not lead to any local symmetry breaking, they are distinguished
by their non-local topologies.

\section{Low Energy Fluid-like State in Non-Symmetry Breaking Phase}

The main goal of condensed matter and many-body physics is to study
the low energy behavior of the system, for the low energy modes responsible
for its core properties. As it is well known that in an ordered symmetric
breaking phase, the Goldstone theorem tells us that there must exist
massless Goldstone modes dominating the low energy behavior. However,
without an analog theorem in a non-symmetry breaking phase and local
order parameter to characterize the phase, low energy modes of a strongly
coupled gauge system is poorly understood and thought notoriously
hard to solve. The observation that topological terms such as Chern-Simons
term in 2+1D are inevitable in a gaped gauge system at finite chemical
potential provides us a route to the problem. The bosonized low energy
effective action Eq.(\ref{eq:abelian-action}) and Eq.(\ref{eq:nonabelian-action})
shows the fact that the leading contribution to the partition function
is the topological term, while the Yang-Mills term is perturbation
due to its strong coupling and the gap $M$. It was first numerically
found \cite{PhysRevLett.59.2095,PhysRevB.39.11879} that the Laughlin
wavefunction borrowed from the theory of Quantum Hall system effectively
minimizes the groundstate energy of the system deviating from the
half-filled large band gap Hubbard model which is exactly a gauge
symmetric system. The bosonized effective theory of the paper is consistent
with the numerical result, and it convinces us that the low energy
modes of a gaped gauge system at finite chemical potential is dominated
by a topological fluid describe by the Chern-Simons theory. The low
energy fluid-like state emerges from the gauge system at finite chemical
potential is in many aspects similar with the quantum Hall fluid $\mathcal{L}=i\frac{\mathrm{Ch}}{4\pi}\epsilon_{\mu\nu\rho}a_{\mu}\partial_{\nu}a_{\rho}$
with Chern number $\mathrm{Ch}=2$ which is well-studied: 

(1) The low energy modes is fluid-like and incompressible, when we
only consider the topological Chern-Simons term. Here the terminology
{}``incompressible'' is tantamount to {}``topological'' since
there is no response to its volume change. When the gaped terms like
the non-topological Yang-Mills term at order $\mathcal{O}(1/M)$ and
$\mathcal{O}(1/g^{2})$ are considered at finite temperature, the
fluid gets its compressibility corrections, thus gradually deviating
from incompressible. As a consequence, the quasi-particle picture
fails, there is no sharp quasi-particle peak in experimental probe. 

(2) The equation of state of the new fluid state of matter deviates
from the idea gas especially at the critical point, since it is purely
topological term dominant at the critical point, only the perturbative
dynamical Yang-Mills term contributes to a quasi-particle/gas -like
behavior.

(3) The fluid-like low energy modes are of conformal at the critical
point due to the fact that the topological action is a fixed point
action.

(4) The gauge charge of the excitation of the phase is fractionalized.

All these salient features of fluid-like low energy modes are expected
shared by the non-Fermi-liquid normal state of high-temperature superconductor,
and quark-gluon plasma of QCD system in finite chemical potential
region beyond ordered phases, just because all these systems are generally
Gribov gaped and strongly coupled gauge symmetric systems as well
as the Chern-Simons term inevitably emerges at finite chemical potential.
Although they are in a non-symmetry breaking phase, they are not completely
disorder as we took it as a matter of course, they are actually topological
ordered, since the system will develops robust groundstate degeneracy
when we put it on base manifolds with different topologies.

\section{Conclusions}

In this paper, due to the fact that gauge systems are gaped by $M$
which sets a scale for the chiral phase transition, we show that the
fermions described by the Grassman numbers can be effectively simulated
by the conventional numbers of bosons in 2+1 dimensions. Based on
the non-locality of fermions, we regard a fermion as an end point
of a bosonic non-local string. This picture introduces an identification
between theories of fermions and bosons. The new bosonization recipe
is systematically developed in 2+1 dimensions. For practical reasons,
we only focus on the massless fermions in this paper. In general,
the new bosons required to bosonize fermions are of a $(d-2)$-form
gauge fields, which explains why a conventional Maxwell/Yang-Mills
type of 1-form gauge bosons is enough to bosonize the fermions in
2+1 dimensions. A more realistic QCD system in 3+1 dimensions requires
a 2-form Kalb-Ramond type gauge fields \cite{PhysRevD.9.2273}, which
is left for future discussions. The feasibility of this identification
between fermions and bosons is reflected by the fact that the fermion
statistic, or equivalently the sign structure of the fermions, can
be totally attributed to the gauge phase controlled by the new gauge
fields.

The bosonization recipe bridges the gap between a gauge system at
finite chemical potential and a pure gauge system. We find that a
Maxwell/Yang-Mills gauge system (with massless fermion) with finite
chemical potential in 2+1 dimensions is equivalent to a gauge system
with new gauge field governed by the action of Chern-Simons form up
to certain gaped corrections. The new pure gauge theories preserves
all symmetry of the original gauge theories at finite chemical potential.
The induced Chern-Simons-Maxwell/Yang-Mills gauge system is well defined
and be our starting point. The new theories provide us a new viewing
angle to the well-known fermion sign problem. The non-trivial phase
of fermion determinant equivalently comes from the Chern-Simons term
of the theories, which could not be eliminated by conventional Wick
rotation due to its metric independent nature. The sign problem can
only be alleviated. Since the gauge fields of the system are gaped
especially at strong coupling, a factorization can be introduced to
hive off the contribution from the Chern-Simons part which is relevant
to the sign problem but irrelevant to the size or temperature dependent
thermodynamic properties.

Since the gauge system at finite chemical potential has identified
to a new pure gauge system, several powerful tools developed in a
pure gauge system can be generalized to the new gauge system. As an
example, the Wilson loop is still valid as an order parameter characterizing
confinement in the new pure gauge system, being a bosonized gauge
system at finite chemical potential. For the Abelian group, the only
effect of chemical potential is found just a shift of the time component
of the gauge field. In the framework, we calculate the chemical potential
dependence of the expectation value of the Wilson loop at strong coupling
limit, and its induced confining potential and transition temperature.
The Wilson's criteria of confinement is generalized to finite chemical
potential. We find that the chemical potential plays the role of weakening
the tension of the confining string, and finally snaps the string
leading to the deconfinement, see Eq.(\ref{eq:area-law-abelian}).
The confined/deconfined phase boundary is obtained in this calculation
Eq.(\ref{eq:phase-boundary-abelian}). Although the chemical potential
is no longer a pure shift to the time component of gauge field in
the non-Abelian case, the analog results can also be derived and generalized
for the non-Abelian gauge system, see the Eq.(\ref{eq:area-law-nonabelian})
and Eq.(\ref{eq:phase-boundary-nonabelian}).

The bosonization is a promising approach to put the chiral symmetry
breaking and the confinement on an equal footing. We find in 2+1 dimensions
that the chiral symmetry breaking is a necessary (not sufficient)
condition for the confinement, so the confined phase is always contained
inside the chiral symmetry breaking phase. As a well known result,
there is only confined phase for the 2+1 dimensional pure gauge system.
When the system is coupled with fermion matter, the system is found
can also be confined, only when the fermions are condensed in such
a way that the gauge symmetry is broken down to the center symmetry.
The deconfined phase can coexist with the chiral symmetry breaking
phase. Different from the chiral symmetry breaking phase of the Landau-Ginzburg
type, the confined and deconfined phase are not characterized by any
local order parameter and/or local symmetry, they are distinguished
by their non-local topological properties.

Beyond the ordered Landau-Ginzburg type phase transition, we study
the low energy modes of a non-symmetry breaking phase at finite chemical
potential. Benefit from the emergence of topological term of a gaped
gauge theories at finite chemical potential, the low energy modes
could be studied analogous with the quantum Hall system. Similar with
the Hall fluid, they are fluid-like without any quasi-particle sharp
peak in spectrum, nearly incompressible, strongly deviate from gas
behavior and gauge charge fractionalized.

The bosonization is a new attempt to study the quantum phase of a
gauge system at finite chemical potential, with a new perspective
to the long-standing fermion sign problem. The method can also be
generalized to (3+1)D \cite{PhysRevB.87.085132,2011arXiv1112.3281D},
based on two main reasons: (1) Non-dynamical components of gauge fields,
e.g. $a_{0}$ (1-form) and $b_{i0}=-b_{0i}$ (2-form), play the roles
of different Lagrange multipliers like different chemical potentials,
which keep corresponding charges conserved. Thus they inevitably lead
to the emergence of topological terms as constraints at finite chemical
potential. (2) The quantum Yang-Mills gauge theory in (3+1)D is generally
believed to be gaped at low energy especially at strong coupling (even
though the exact proof is still lacking). The gap strongly suppresses
the dynamical infrared modes induced from Yang-Mills term and protects
the emergent topological degrees of freedom at low energy, which will
prohibits a hydrodynamic low energy behavior in gauge systems. It
suggests that the fluid-like quark-gluon plasma found at Relativistic
Heavy Ion Collider (RHIC) may has deep connection to the topological
fluid governing by the topological terms (e.g. the BF-terms and theta-terms
in (3+1)D). We believe that bosonization method is worth pursuing
and hopeful to give us new insights to the physics at finite chemical
potential.
\begin{acknowledgments}
This work was supported in part by the National Science Foundation
of China (NSFC) under Grant No.11205149.
\end{acknowledgments}

\section*{Appendix I}

In this appendix we proof that the bosonization recipe reproduces
the fermion statistic. By using Eq.(\ref{eq:bosonize}). The equal
time anti-commutation relation for the fermions is given by

\begin{align}
\psi(\vec{x}_{1},\tau)\psi(\vec{x}_{2},\tau) & =\lim_{\vec{x}_{f}\rightarrow-\infty}\exp\left(-i\int_{\vec{x}_{f}}^{\vec{x}_{1}}dy^{i}a_{i}(\vec{y},\tau)\right)\exp\left(-i\int_{\vec{x}_{f}}^{\vec{x}_{2}}dy^{i}a_{i}(\vec{y},\tau)\right)\nonumber \\
 & =\lim_{\vec{x}_{f}\rightarrow-\infty}e^{i\Phi^{\prime}(\vec{x}_{1},\vec{x}_{2})}\exp\left(-i\int_{\vec{x}_{f}}^{\vec{x}_{2}}dy^{i}a_{i}(\vec{y},\tau)\right)\exp\left(-i\int_{\vec{x}_{f}}^{\vec{x}_{1}}dy^{i}a_{i}(\vec{y},\tau)\right)\nonumber \\
 & =e^{i\Phi(\vec{x}_{1},\vec{x}_{2})}\psi(\vec{x}_{2},\tau)\psi(\vec{x}_{1},\tau),\end{align}
in which we have used the identity\begin{equation}
e^{A}e^{B}\equiv e^{B}e^{A}e^{-[A,B]}.\end{equation}
The canonical commutation relation of $a_{i}$ field is written as\begin{align}
i\delta^{2}(\vec{x}-\vec{y}) & =\left[a_{i}(\vec{x},\tau),\Pi_{i}(\vec{y},\tau)\right]=\left[a_{i}(\vec{x},\tau),\frac{\delta S}{\delta\partial_{0}a_{i}(\vec{y},\tau)}\right]\nonumber \\
 & =\pm\frac{1}{\pi}\epsilon_{ij}[a_{i}(\vec{x},\tau),a_{j}(\vec{y},\tau)],\end{align}
in which the conjugate momentum is calculated by using the action
Eq.(\ref{eq:abelian-action}) or Eq.(\ref{eq:nonabelian-action})
in which we have neglected the Maxwell/Yang-Mills terms suppressed
by gap and strong coupling. So the equal time anti-commutation relation
of fermions can be given as follows,

\begin{align}
i\Phi(\vec{x}_{1},\vec{x}_{2}) & =\lim_{\vec{x}_{f}\rightarrow-\infty}\left[\int_{\vec{x}_{f}}^{\vec{x}_{1}}dy_{1}^{i}a_{i}(\vec{y}_{1},\tau),\int_{\vec{x}_{f}}^{\vec{x}_{2}}dy_{2}^{j},a_{j}(\vec{y}_{2},\tau)\right]\nonumber \\
 & =\lim_{\vec{x}_{f}\rightarrow-\infty}\int_{\vec{x}_{f}}^{\vec{x}_{1}}dy_{1}^{i}\int_{\vec{x}_{f}}^{\vec{x}_{2}}dy_{2}^{j}\left[a_{i}(\vec{y}_{1},\tau),a_{j}(\vec{y}_{2},\tau)\right]\nonumber \\
 & =\pm i\pi\lim_{\vec{x}_{f}\rightarrow-\infty}\int_{\vec{x}_{f}}^{\vec{x}_{1}}dy_{1}^{i}\int_{\vec{x}_{f}}^{\vec{x}_{2}}dy_{2}^{j}\epsilon_{ij}\delta^{2}(\vec{y}_{1}-\vec{y}_{2}).\end{align}

Note that the integral of the delta function equals 1 when the coordinates
of two paths meet at the fixed point $\vec{x}_{f}$, so we reproduce
the equal time anti-commutation relation of fermion $i\Phi(\vec{x}_{1},\vec{x}_{2})=\pm i\pi$.
It is worth mentioning that the $\psi$ satisfies fermionic statistics
on the equal time spatial hyper-plane, independent with any specific
detail of the paths, so it is gauge independent. The arbitrariness
of the plus and minus sign here is crucial for the parity conservation
for the action. The deduction shows how the anti-commutation relation
of fermions connects with the commutator of the $a$ fields, and its
relation with the coefficient of the Chern-Simons action.

\section*{Appendix II}

In this appendix, we proof the bosonized form of the current in Abelian
and non-Abelian case in the dictionaries. First, we proof the bosonized
current for the Abelian case. The definition of current is\begin{align}
J_{i}(x) & =:\bar{\psi}(x)\gamma_{i}\psi(x):\nonumber \\
 & =\lim_{\epsilon\rightarrow0}:\psi^{\dagger}(x+\epsilon)\gamma_{0}\gamma_{i}\psi(x):,\end{align}
in which $::$ standards for the normal ordering, so we use the identity
between the normal ordering and time ordering\begin{equation}
:\phi(x)\chi(y):=T\phi(x)\chi(y)-\left\langle T\phi(x)\chi(y)\right\rangle ,\end{equation}
we get\begin{align}
J_{i}(x) & =\lim_{\epsilon\rightarrow0}\left[T\psi^{\dagger}(x+\epsilon)\gamma_{0}\gamma_{i}\psi(x)-\left\langle \psi^{\dagger}(x+\epsilon)\gamma_{0}\gamma_{i}\psi(x)\right\rangle \right].\end{align}
For the massless fermions, there are two chiral components $\psi_{\pm}$.
We set $u_{\pm}$ the spinor bases with unit norm for different chirality.
Then we write the current as

\begin{align}
J_{i}(x) & =\lim_{\epsilon\rightarrow0}\left[T\psi^{\dagger}(x+\epsilon)\gamma_{0}\gamma_{i}\psi(x)-\left\langle \psi^{\dagger}(x+\epsilon)\gamma_{0}\gamma_{i}\psi(x)\right\rangle \right]\nonumber \\
 & =\lim_{\epsilon\rightarrow0}\left[T\hat{\psi}^{\dagger}(x+\epsilon)\hat{\psi}(x)-\left\langle \hat{\psi}^{\dagger}(x+\epsilon)\hat{\psi}(x)\right\rangle \right]\left(u_{\pm}^{\dagger}\gamma_{0}\gamma_{i}u_{\pm}\right)\nonumber \\
 & =\lim_{\epsilon\rightarrow0}\left[:\exp\left(i\int_{-\infty}^{x+\epsilon}dy^{j}a_{j}(y)\right)::\exp\left(-i\int_{-\infty}^{x}dy^{j}a_{j}(y)\right):-\frac{1}{2\pi\epsilon^{2}}\right]\left(u_{\pm}^{\dagger}\gamma_{0}\gamma_{i}u_{\pm}\right),\end{align}
in which we have used the 2+1 dimensional correlator of fermions in
the bases of $u_{\pm}$,\begin{align}
\left\langle \hat{\psi}^{\dagger}(x+\epsilon)\hat{\psi}(x)\right\rangle  & =\int\frac{1}{(2\pi)^{2}}d^{2}k\int\frac{1}{(2\pi)^{2}}d^{2}q\langle0|\alpha^{\dagger}(k)\alpha(q)|0\rangle e^{i\vec{k}\cdot\vec{x}-i\vec{q}\cdot\vec{x}^{\prime}}\nonumber \\
 & =\frac{1}{2\pi}\int_{k>0}dkke^{i\vec{k}\cdot(\vec{x}-\vec{x}^{\prime})}\nonumber \\
 & =\frac{1}{2\pi}\frac{1}{\left(\vec{x}-\vec{x}^{\prime}\right)^{2}}.\end{align}
By using the identity of normal ordering\begin{equation}
:e^{ia\phi(z)}::e^{ib\phi(z^{\prime})}:=:e^{ia\phi(z)+ib\phi(z^{\prime})}:e^{-ab\left\langle \phi(z)\phi(z^{\prime})\right\rangle },\end{equation}
we have\begin{align}
J_{i} & =\lim_{\epsilon\rightarrow0}\left[\exp\left(i\int_{-\infty}^{x+\epsilon}dy^{j}a_{j}(y)-i\int_{-\infty}^{x}dy^{j}a_{j}(y)\right)\exp\left(\int_{-\infty}^{x+\epsilon}dy_{1}^{j}\int_{-\infty}^{x}dy_{2}^{k}\left\langle a_{j}(y_{1}),a_{k}(y_{2})\right\rangle \right)-\frac{1}{2\pi\epsilon^{2}}\right]\left(u_{\pm}^{\dagger}\gamma_{0}\gamma_{i}u_{\pm}\right)\nonumber \\
 & =\lim_{\epsilon\rightarrow0}\left[\exp\left(i\int_{x}^{x+\epsilon}dy^{j}a_{j}(y)\right)\frac{1}{2\pi\epsilon^{2}}-\frac{1}{2\pi\epsilon^{2}}\right]\left(u_{\pm}^{\dagger}\gamma_{0}\gamma_{i}u_{\pm}\right),\end{align}
in which we have used the correlator of $a$\begin{equation}
\langle a_{i}(x)a_{j}(x^{\prime})\rangle=\frac{1}{2\pi}\epsilon_{ijk}\frac{x_{k}-x_{k}^{\prime}}{\left|x-x^{\prime}\right|^{3}}.\end{equation}
Then we use the Stokes's theorem,\begin{align}
J_{i} & =\lim_{\epsilon_{j},\epsilon_{k}\rightarrow0}\left[\exp\left(i\sum_{j,k\neq i}\int_{x}^{x+\epsilon_{j}}\int_{x}^{x+\epsilon_{k}}dy^{j}dy^{k}\epsilon_{jk}\partial_{j}a_{k}\right)\frac{1}{2\pi\epsilon^{2}}-\frac{1}{2\pi\epsilon^{2}}\right]\left(u_{\pm}^{\dagger}\gamma_{0}\gamma_{i}u_{\pm}\right)\nonumber \\
 & =\lim_{\epsilon_{j},\epsilon_{k}\rightarrow0}\left[\exp\left(i\sum_{j,k\neq i}\epsilon_{jk}\partial_{j}a_{k}\epsilon_{j}\epsilon_{k}\right)\frac{1}{2\pi\epsilon^{2}}-\frac{1}{2\pi\epsilon^{2}}\right]\left(u_{\pm}^{\dagger}\gamma_{0}\gamma_{i}u_{\pm}\right)\nonumber \\
 & =\lim_{\epsilon\rightarrow0}\left[\left(1+i\sum_{j,k\neq i}\epsilon_{jk}\partial_{j}a_{k}\epsilon^{2}+...\right)\frac{1}{2\pi\epsilon^{2}}-\frac{1}{2\pi\epsilon^{2}}\right]\left(u_{\pm}^{\dagger}\gamma_{0}\gamma_{i}u_{\pm}\right)\nonumber \\
 & =\frac{1}{2\pi}i\epsilon_{jk}\left(u_{\pm}^{\dagger}\gamma_{0}\gamma_{i}u_{\pm}\right)\partial_{j}a_{k}\nonumber \\
 & =\pm\frac{1}{2\pi}i\epsilon_{ijk}\partial_{j}a_{k},\end{align}
in which we have used the identities\begin{align}
\epsilon_{ij}\gamma_{0} & =\sum_{k\neq i,j}\epsilon_{ijk}\gamma_{k}.\end{align}

For the non-Abelian case, the current is defined as\begin{align}
J_{i}^{I}(x) & =\lim_{\epsilon\rightarrow0}\mathrm{tr}\left[:\bar{\psi}(x+\epsilon)\gamma_{i}t^{I}\psi(x):\right]\nonumber \\
 & =\mathrm{tr}\left[T\psi^{\dagger}(x+\epsilon)\gamma_{0}\gamma_{i}t^{I}\psi(x)-\left\langle T\psi(x+\epsilon)\gamma_{0}\gamma_{i}t^{I}\psi(x)\right\rangle \right].\end{align}
We write the current in its eigenstates of $t^{I}$ and $u_{\pm}$,\begin{equation}
J_{i}^{I}=\mathrm{tr}\lim_{\epsilon\rightarrow0}\left[\exp\left(i\int_{-\infty}^{x+\epsilon}dy^{j}a_{j}^{J}t^{J}\right)\exp\left(-i\int_{-\infty}^{x}dy^{j}a_{j}^{J}t^{J}\right)-\frac{1}{2\pi\epsilon^{2}}\right]\left(U^{\dagger}t^{I}U\right)\left(u_{\pm}^{\dagger}\gamma_{0}\gamma_{i}u_{\pm}\right)\end{equation}
where $U$ is unit norm eigen-vector of $t^{I}$. By using the identity

\begin{equation}
e^{A}e^{B}=e^{A+B+\frac{1}{2}[A,B]},\quad:e^{ia\phi(z)}::e^{ib\phi(z^{\prime})}:=:e^{ia\phi(z)+ib\phi(z^{\prime})}:e^{-ab\left\langle \phi(z)\phi(z^{\prime})\right\rangle },\label{eq:identity-normal-order}\end{equation}
we get\begin{align}
J_{i}^{I} & =\mathrm{tr}\lim_{\epsilon\rightarrow0}\left[\exp\left(i\int_{-\infty}^{x+\epsilon}dy^{j}a_{j}^{J}t^{J}-i\int_{-\infty}^{x}dy^{j}a_{j}^{J}t^{J}+\frac{1}{2}\left[\int_{-\infty}^{x+\epsilon}dy^{j}a_{j}^{K}t^{K},\int_{-\infty}^{x}dy^{k}a_{k}^{L}t^{L}\right]\right)\times\right.\nonumber \\
 & \left.\qquad\times\exp\left(\int_{-\infty}^{x+\epsilon}dy_{1}^{j}\int_{-\infty}^{x}dy_{2}^{k}\left\langle a_{j}^{K}(y_{1})t^{K},a_{k}^{L}(y_{2})t^{L}\right\rangle \right)-\frac{1}{2\pi\epsilon^{2}}\right]\left(U^{\dagger}t^{I}U\right)\left(u_{\pm}^{\dagger}\gamma_{0}\gamma_{i}u_{\pm}\right)\nonumber \\
 & =\mathrm{tr}\lim_{\epsilon\rightarrow0}\left[\exp\left(i\int_{x}^{x+\epsilon}dy^{j}a_{j}^{J}t^{J}+\int_{x}^{x+\epsilon_{j}}\int_{x}^{x+\epsilon_{k}}dy^{j}dy^{k}[a_{j}^{K}t^{K},a_{k}^{L}t^{L}]\right)\times\right.\nonumber \\
 & \left.\qquad\times\exp\left(\int_{-\infty}^{x+\epsilon}dy_{1}^{j}\int_{-\infty}^{x}dy_{2}^{k}\left\langle a_{j}^{K}(y_{1})t^{K},a_{k}^{L}(y_{2})t^{L}\right\rangle \right)-\frac{1}{2\pi\epsilon^{2}}\right]\left(U^{\dagger}t^{I}U\right)\left(u_{\pm}^{\dagger}\gamma_{0}\gamma_{i}u_{\pm}\right),\end{align}
in which we have used the following relation\begin{equation}
\frac{1}{2}\left[\int_{-\infty}^{x+\epsilon}dy^{j}a_{j},\;\int_{-\infty}^{x+\epsilon}dy^{k}a_{k}\right]=\int_{x}^{x+\epsilon_{j}}dy^{j}\int_{x}^{x+\epsilon_{k}}dy^{k}[a_{j},a_{k}].\label{eq:identity-integral-commutator}\end{equation}
Then using the Stokes's theorem\begin{align}
J_{i}^{I} & =\mathrm{tr}\lim_{\epsilon_{j},\epsilon_{k}\rightarrow0}\left[\exp\left(i\sum_{j,k\neq i}\int_{x}^{x+\epsilon_{j}}\int_{x}^{x+\epsilon_{k}}dy^{j}dy^{k}\epsilon_{jk}\partial_{j}a_{k}^{J}t^{J}-ia_{j}^{K}a_{k}^{L}[t^{K},t^{L}]\right)\frac{1}{2\pi\epsilon^{2}}-\frac{1}{2\pi\epsilon^{2}}\right]\left(U^{\dagger}t^{I}U\right)\left(u_{\pm}^{\dagger}\gamma_{0}\gamma_{i}u_{\pm}\right)\nonumber \\
 & =\mathrm{tr}\lim_{\epsilon_{j},\epsilon_{k}\rightarrow0}\left[\exp\left[\sum_{j,k\neq i}\epsilon_{j}\epsilon_{k}i\epsilon_{jk}\left(\partial_{j}a_{k}^{J}t^{J}+\theta^{JKL}a_{j}^{K}a_{k}^{L}t^{J}\right)\right]\frac{1}{2\pi\epsilon^{2}}-\frac{1}{2\pi\epsilon^{2}}\right]\left(U^{\dagger}t^{I}U\right)\left(u_{\pm}^{\dagger}\gamma_{0}\gamma_{i}u_{\pm}\right)\nonumber \\
 & =\mathrm{tr}\lim_{\epsilon\rightarrow0}\left[\left[1+\epsilon^{2}i\sum_{j,k\neq i}\epsilon_{jk}\left(\partial_{j}a_{k}^{J}t^{J}+\theta^{JKL}a_{j}^{K}a_{k}^{L}t^{J}\right)+...\right]\frac{1}{2\pi\epsilon^{2}}-\frac{1}{2\pi\epsilon^{2}}\right]\left(U^{\dagger}t^{I}U\right)\left(u_{\pm}^{\dagger}\gamma_{0}\gamma_{i}u_{\pm}\right)\nonumber \\
 & =\frac{1}{2\pi}i\epsilon_{jk}\left(u_{\pm}^{\dagger}\gamma_{0}\gamma_{i}u_{\pm}\right)\mathrm{tr}\left(\partial_{j}a_{k}^{J}+\theta^{JKL}a_{j}^{K}a_{k}^{L}\right)\mathrm{tr}\left(t^{J}U^{\dagger}t^{I}U\right)\nonumber \\
 & =\pm\frac{1}{2\pi}i\epsilon_{ijk}\mathrm{tr}\left(\partial_{j}a_{k}^{I}+\theta^{IJK}a_{j}^{J}a_{k}^{K}\right).\end{align}

\section*{Appendix III}

In this appendix we proof the bosonized form of kinetic term in the
dictionaries. The massless fermion kinetic energy is identified to
the Chern-Simons term at the Lagrangian (classical) level. Using the
Eq.(\ref{eq:bosonize}), the kinetic energy is given by\begin{align}
i\bar{\psi}\gamma_{i}\partial_{i}\psi & =i\bar{\psi}\gamma_{i}\partial_{i}\exp\left(-i\int_{-\infty}^{x}dy^{i}a_{i}\right)\nonumber \\
 & =i\bar{\psi}\gamma_{i}\exp\left(-i\int_{-\infty}^{x}dy^{i}a_{i}\right)\partial_{i}\left(-i\int_{-\infty}^{x}dy^{j}a_{j}\right)\nonumber \\
 & =\bar{\psi}\gamma_{i}\psi a_{i}.\end{align}
By using the bosonized form of current deduced from the appendix II,\begin{equation}
J_{i}=\bar{\psi}\gamma_{i}\psi=\pm\frac{1}{2\pi}i\epsilon_{ijk}\partial_{j}a_{k},\end{equation}
so\begin{equation}
i\bar{\psi}\gamma_{i}\partial_{i}\psi=J_{i}a_{i}=\pm\frac{1}{2\pi}i\epsilon_{ijk}a_{i}\partial_{j}a_{k}.\end{equation}
The non-Abelian counterpart is\begin{align}
i\bar{\psi}\gamma_{i}\partial_{i}\psi & =i\bar{\psi}\gamma_{i}\partial_{i}\exp\left(-i\int_{-\infty}^{x}dy^{i}a_{i}^{I}t^{I}\right)\nonumber \\
 & =i\bar{\psi}\gamma_{i}\exp\left(-i\int_{-\infty}^{x}dy^{i}a_{i}^{J}t^{J}\right)\partial_{i}\left(-i\int_{-\infty}^{x}dy^{i}a_{i}^{I}t^{I}\right)\nonumber \\
 & =\bar{\psi}\gamma_{i}\exp\left(-i\int_{-\infty}^{x}dy^{l}a_{l}^{J}t^{J}\right)\left(a_{i}^{I}t^{I}\right).\end{align}
By using the relation\begin{equation}
\exp\left(-i\alpha^{J}t^{J}\right)\left(a_{i}^{I}t^{I}\right)\exp\left(i\alpha^{J}t^{J}\right)=a_{i}^{I}t^{I}+i\left[a_{i}^{I}t^{I},\alpha^{J}t^{J}\right]\end{equation}
we have\begin{align}
i\bar{\psi}\gamma_{i}\partial_{i}\psi & =\bar{\psi}\gamma_{i}\left(a_{i}^{I}t^{I}+i\int_{-\infty}^{x}dy^{l}\left[a_{i}^{I}t^{I},a_{l}^{J}t^{J}\right]\right)\exp\left(-i\int_{-\infty}^{x}dy^{j}a_{j}^{J}t^{J}\right)\nonumber \\
 & =\mathrm{tr}a_{i}^{I}J_{i}^{I}+\mathrm{tr}\int_{-\infty}^{x}dy^{l}\left(\bar{\psi}\gamma_{i}ia_{i}^{I}a_{l}^{J}\left[t^{I},t^{J}\right]\psi\right).\end{align}
We use the non-Abelian form of the current\begin{equation}
J_{i}^{I}=\bar{\psi}\gamma_{i}t^{I}\psi=\pm\frac{1}{2\pi}i\epsilon_{ijk}\mathrm{tr}\left(\partial_{j}a_{k}^{I}+\theta^{IJK}a_{j}^{J}a_{k}^{K}\right),\end{equation}
then\begin{align}
i\bar{\psi}\gamma_{i}\partial_{i}\psi & =\mathrm{tr}a_{i}^{I}J_{i}^{I}-\theta^{IJK}\mathrm{tr}\int_{-\infty}^{x}dy^{l}\left(a_{i}^{I}a_{l}^{J}J_{i}^{K}\right)\nonumber \\
 & =\mathrm{tr}a_{i}^{I}J_{i}^{I}\mp\frac{1}{2\pi}i\epsilon_{ijk}\theta^{IJK}\mathrm{tr}\int_{-\infty}^{x}dy^{l}\left[a_{i}^{I}a_{l}^{J}\left(\partial_{j}a_{k}^{K}+\theta^{KAB}a_{j}^{A}a_{k}^{B}\right)\right].\end{align}
Since $\mathrm{tr}a^{4}=0$, so\begin{align}
i\bar{\psi}\gamma_{i}\partial_{i}\psi & =\mathrm{tr}a_{i}^{I}J_{i}^{I}\mp\frac{1}{2\pi}i\epsilon_{ijk}\theta^{IJK}\mathrm{tr}\int_{-\infty}^{x}dy^{l}\left[a_{i}^{I}a_{l}^{J}\left(\partial_{j}a_{k}^{K}\right)\right]\nonumber \\
 & =\mathrm{tr}a_{i}^{I}J_{i}^{I}\mp\frac{1}{2\pi}i\epsilon_{ijk}\theta^{IJK}\frac{1}{3}\mathrm{tr}\int_{-\infty}^{x}dy^{l}\left[\partial_{j}\left(a_{i}^{I}a_{l}^{J}a_{k}^{K}\right)\right]\nonumber \\
 & =\pm\frac{1}{2\pi}i\epsilon_{ijk}\mathrm{tr}a_{i}^{I}\left(\partial_{j}a_{k}^{I}+\theta^{IJK}a_{j}^{J}a_{k}^{K}\right)\mp\frac{1}{2\pi}i\epsilon_{ijk}\theta^{IJK}\frac{1}{3}\mathrm{tr}a_{i}^{I}a_{j}^{J}a_{k}^{K}\nonumber \\
 & =\pm\frac{1}{2\pi}i\epsilon_{ijk}\mathrm{tr}a_{i}^{I}\left(\partial_{j}a_{k}^{I}+\frac{2}{3}\theta^{IJK}a_{j}^{J}a_{k}^{K}\right).\end{align}

\section*{Appendix IV}

In this appendix we proof the bosonized form of the chiral densities.
For the massless fermions, there are two chiral components, $\psi_{\pm}$.
The definition of the chiral densities are given by\begin{equation}
\bar{\psi}_{+}\psi_{-}=\bar{\psi}\left(1+\gamma_{5}\right)\psi,\quad\bar{\psi}_{-}\psi_{+}=\bar{\psi}\left(1-\gamma_{5}\right)\psi,\end{equation}
where\begin{align}
\bar{\psi}_{-}(x)\psi_{+}(x) & =\lim_{\epsilon\rightarrow0}=\bar{\psi}_{-}(x+\epsilon)\psi_{+}(x)\nonumber \\
 & =\lim_{\epsilon\rightarrow0}:\exp\left(i\int_{-\infty}^{x+\epsilon}dy^{i}\bar{a}_{i}\right)::\exp\left(-i\int_{-\infty}^{x}dy^{i}a_{i}\right):\end{align}
where $a_{i}=a_{i}^{I}t^{I}$, $a=a_{(0)}+\partial a+...$ is the
left mover and $\bar{a}=a_{(0)}-\partial a+...$ is the right mover.
By using the identity Eq.(\ref{eq:identity-normal-order}) and Eq.(\ref{eq:identity-integral-commutator}),
we have\begin{align}
\bar{\psi}_{-}(x)\psi_{+}(x) & =\lim_{\epsilon\rightarrow0}\mathrm{tr}\exp\left(i\int_{-\infty}^{x+\epsilon}dy^{i}\bar{a}_{i}-i\int_{-\infty}^{x}dy^{i}a_{i}+\frac{1}{2}\left[\int_{-\infty}^{x+\epsilon}dy^{i}\bar{a}_{i},\int_{-\infty}^{x}dy^{i}a_{i}\right]\right)\exp\left(\int_{-\infty}^{x+\epsilon}\int_{-\infty}^{x}dy^{i}dy^{j}\left\langle \bar{a}_{i}a_{j}\right\rangle \right)\nonumber \\
 & =\lim_{\epsilon\rightarrow0}M^{2}\mathrm{tr}\exp\left(i\int_{-\infty}^{x+\epsilon}dy^{i}\bar{a}_{i}-i\int_{-\infty}^{x}dy^{i}a_{i}+\int_{x}^{x+\epsilon_{i}}\int_{-\infty}^{x+\epsilon_{j}}dy^{i}dy^{j}[\bar{a}_{i},a_{j}]\right)\nonumber \\
 & =\lim_{\epsilon\rightarrow0}M^{2}\mathrm{tr}\exp\left(-i\int_{x}^{x+\epsilon}dy^{i}a_{i}+\int_{x}^{x+\epsilon_{i}}\int_{-\infty}^{x+\epsilon_{j}}dy^{i}dy^{j}[\bar{a}_{i},a_{j}]\right)\nonumber \\
 & =\lim_{\epsilon\rightarrow0}M^{2}\mathrm{tr}\exp\left(-i\int_{x}^{x+\epsilon_{i}}\int_{x}^{x+\epsilon_{j}}dy^{i}dy^{j}\epsilon_{ij}\partial_{i}a_{j}+i[a_{i},a_{j}]\right)\nonumber \\
 & =\lim_{\Sigma_{x}\rightarrow0}M^{2}\mathrm{tr}\exp\left(-i\iint_{\Sigma_{x}=\epsilon_{i}\times\epsilon_{j}}dy^{i}dy^{j}f_{ij}\right)\nonumber \\
 & =M^{2}\mathrm{tr}\exp\left(-i\Phi(x)\right),\end{align}
where\begin{equation}
M^{2}(\epsilon)=\exp\left(\int_{-\infty}^{x+\epsilon}\int_{-\infty}^{x}dy^{i}dy^{j}\left\langle a_{i}(-k)a_{j}(k)\right\rangle \right)=\frac{1}{2\pi\epsilon^{2}}\end{equation}
is a universal cut-off scale with dimensions of mass squared, the
cut-off $\epsilon$ is the shortest distance between the fermions.
Similarly, one find\begin{align}
\bar{\psi}_{+}(x)\psi_{-}(x) & =\lim_{\epsilon\rightarrow0}\mathrm{tr}\exp\left(i\int_{-\infty}^{x+\epsilon}dy^{i}a_{i}-i\int_{-\infty}^{x}dy^{i}\bar{a}_{i}+\frac{1}{2}\left[\int_{-\infty}^{x+\epsilon}dy^{i}a_{i},\int_{-\infty}^{x}dy^{i}\bar{a}_{i}\right]\right)\exp\left(\int_{-\infty}^{x+\epsilon}\int_{-\infty}^{x}dy^{i}dy^{j}\left\langle a_{i}\bar{a}_{j}\right\rangle \right)\nonumber \\
 & =\lim_{\epsilon\rightarrow0}M^{2}\mathrm{tr}\exp\left(i\int_{-\infty}^{x+\epsilon}dy^{i}a_{i}-i\int_{-\infty}^{x}dy^{i}\bar{a}_{i}+\int_{x}^{x+\epsilon_{i}}\int_{-\infty}^{x+\epsilon_{j}}dy^{i}dy^{j}[a_{i},\bar{a}_{j}]\right)\nonumber \\
 & =\lim_{\epsilon\rightarrow0}M^{2}\mathrm{tr}\exp\left(i\int_{x}^{x+\epsilon}dy^{i}a_{i}+\int_{x}^{x+\epsilon_{i}}\int_{-\infty}^{x+\epsilon_{j}}dy^{i}dy^{j}[a_{i},\bar{a}_{j}]\right)\nonumber \\
 & =\lim_{\epsilon\rightarrow0}M^{2}\mathrm{tr}\exp\left(i\int_{x}^{x+\epsilon_{i}}\int_{x}^{x+\epsilon_{j}}dy^{i}dy^{j}\epsilon_{ij}\partial_{i}a_{j}+i[a_{i},a_{j}]\right)\nonumber \\
 & =\lim_{\Sigma_{x}\rightarrow0}M^{2}\mathrm{tr}\exp\left(i\iint_{\Sigma_{x}=\epsilon_{i}\times\epsilon_{j}}dy^{i}dy^{j}f_{ij}\right)\nonumber \\
 & =M^{2}\mathrm{tr}\exp\left(i\Phi(x)\right),\end{align}
where $\Phi$ is the flux density at $x$. This result includes both
the Abelian and non-Abelian case, the bracket $[a_{i},a_{j}]$ automatically
vanishes in the Abelian case. This result translates the (left/right)
chiral density to the flux/anti-flux vortex density in the dictionaries
\begin{equation}
\bar{\psi}\left(1\pm\gamma_{5}\right)\psi=M^{2}\mathrm{tr}\exp\left(\pm i\Phi(x)\right).\end{equation}
The result shows the scalar and axial vector is bosonized as\begin{align}
\bar{\psi}(x)\psi(x) & =\frac{1}{2}\left(\bar{\psi}_{+}\psi_{-}+\bar{\psi}_{-}\psi_{+}\right)=\frac{1}{2}M^{2}\left(\mathrm{tr}e^{i\Phi}+\mathrm{tr}e^{-i\Phi}\right)=M^{2}\mathrm{tr}\cos\left(\Phi\right),\nonumber \\
\bar{\psi}i\gamma_{5}\psi & =M^{2}\mathrm{tr}\sin\left(\Phi\right).\end{align}
So the mass term is written as\begin{equation}
m\bar{\psi}\psi=mM^{2}\mathrm{tr}\cos\left(\Phi\right).\end{equation}

\bibliographystyle{apsrev}

\end{document}